\documentclass[12pt]{article}
\usepackage{amsmath}
\usepackage{amssymb}
\usepackage[russian]{babel}
\usepackage{algpseudocode}
\usepackage[left=2cm,right=2cm,
    top=2cm,bottom=2cm,bindingoffset=0cm]{geometry}

\usepackage{graphicx}
\graphicspath{{pictures/}}
\DeclareGraphicsExtensions{.pdf,.png,.jpg}

\newtheorem{note}{Замечание}
\newtheorem{mydef}{Определение}
\newtheorem{lemma}{Лемма}
\newtheorem{theorem}{Теорема}
\newtheorem{statement}{Утверждение}

\title{An Exponential Envy-Free Cake Cutting Protocol for $n$ Agents}
\author{Georgy Sokolov}
\date{}
\begin{document}
	\maketitle
	\vspace{20 pt}

\begin{abstract}
	Мы рассматриваем задачу о дележе пирога без зависти на $n$ участников. Первый ограниченный по числу запросов протокол был предложен Азизом и Маккензи в 2016 году \cite{AzizMakkenzi}. Недостатком этого протокола является его высокая сложность, в \cite{AzizMakkenzi} сложность протокола оценивается $n^{n^{n^{n^{n^n}}}}$ запросами. Также авторы протокола заявляли, что сложность можно уменьшить до $n^{n^n}$ запросов \cite{QuantaMagazin}. Другим недостатком протокола является высокая сложность его понимания. Мы внесём небольшие изменения в протокол Азиза-Маккензи, упростим его изложение, более аккуратно оценим требуемое количество операций и в результате получим алгоритм, использующий не более, чем $n^{8n^2(1 + o(1))}$ операций.
\end{abstract}

	\section{Постановка задачи}
	Делёж пирога "--- метафора, описывающая большое количество практических задач, заключающихся в справедливом дележе какого-то разнородного непрерывного блага между несколькими участниками дележа, имеющими разные предпочтения. Пусть $N = \{1, 2, ..., n\}$ "--- множество участников дележа. Пирог представляется в виде отрезка $[0, 1]$, кусок пирога "--- объединение конечного числа непересекающихся отрезков. Требуется разделить пирог между участниками дележа, то есть представить его в виде объединения $n$ кусков, которые пересекаются только по границам отрезков, из которых состоят куски, и дать каждому участнику дележа по одному куску. При этом у каждого участника есть неизвестная нам функция предпочтений $v_i \, (i \in N)$, определённая на конечных объединениях отрезков. Мы предполагаем, что все функции предпочтений удовлетворяют следующим условиям:
\begin{itemize}
  \item для любого $A$ выполнено $v_i(A) \geq 0$ (неотрицательность);
  \item $v_i(A) + v_i(B) = v_i(A \cup B)$, если $A$ и $B$ не пересекаются или $A \cap B$ состоит из конечного числа точек (аддитивность);
  \item Для любого отрезка $[a, b]$ и для любого числа $\lambda \in [0, 1]$ найдётся такое $c \in [a, b]$, что $v_i([a, c]) = \lambda \cdot v_i([a, b])$ (непрерывность). Из непрерывности следует, что для любого $a \in [0, 1]$ выполнено $v_i([a, a]) = 0$.
\end{itemize}
Будем обозначать делёж, в котором $i$-й участник получил кусок $S_i$, как $S = (S_1, S_2, ..., S_n)$. Обычно в задаче о дележе без зависти к дележу предъявляется одно из двух требований:
\begin{itemize}
  \item Пропорциональность: $\forall i \in N \, v_i(S_i) \geq \frac{1}{n} v_i([0, 1])$;
  \item Отсутствие зависти: $\forall i \in N \ \forall j \in N \, v_i(S_i) \geq v_i(S_j)$.
\end{itemize}
Очевидно, что из отсутствия зависти следует пропорциональность. Мы будем искать делёж, для которого выполнено отсутствие зависти.
Будем называть $S$ частичным дележом, если $\bigcup^{n}_{i=i}S_i \neq [0, 1]$, то есть осталась неразделённая часть пирога $R = [0, 1] \setminus \bigcup^{n}_{i=i}S_i$. Для частичного дележа отсутсвие зависти определяется так же, как и для дележа всего пирога. Заметим, что если частично разделить пирог без зависти, а потом разделить без зависти остаток, получится делёж без зависти. Действительно, каждый участник будет считать, что он получил не меньше любого другого участника в каждой из двух частей дележа, поэтому из-за аддитивности функции предпочтений он будет считать, что и во всём дележе он получил не меньше любого другого участника дележа. Если $S = (S_1, S_2, \ldots, S_n)$ "--- частичный делёж, $R$ "--- остаток при дележе $S$, $S' = (S'_1, S'_2, \ldots, S'_n)$ "--- частичный делёж $R$ с остатком $R'$, то будем обозначать частичный делёж $(S_1 \cup S'_1, S_2 \cup S'_2, \ldots, S_n \cup S'_n)$ с остатком $R'$ через $S \cup S'$.

	\subsection{Модель Робертсона--Уэбба}
Стандартной моделью для измерения сложности протоколов справедливого дележа является модель Робертсона-Уэбба. В этой модели есть два типа запросов:
\begin{itemize}
  \item Запрос разрезания: по участнику дележа $i$, точке $x$ и действительному числу $r$ таким, что $v_i([x, 1]) \geq r$, получаем такую точку $y$, что $v_i([x, y]) = r$ (просим $i$ отрезать кусок, равный $r$ с началом в $x$).
  \item Запрос измерения: по участнику дележа $i$ и отрезку $[x, y]$ получаем действительное число $v_i([x, y])$.
\end{itemize}
Заметим, что для практики более важной характеристикой является количество запросов разрезания (сколько разрезов надо сделать при дележе пирога). Следующее замечание показывает, что, рассматривая только запросы разрезания, мы не сильно уменьшим общее число запросов.

\begin{note}
Во всех известных алгоритмах дележа без зависти запросы измерения используются только для отрезков с концами в точках 0, 1 и точках из запросов разрезания. Поэтому, если протокол использует $k$ запросов разрезания, то можно сделать так, что он будет использовать не более, чем $(n - 1) k + n$ запросов измерения. Для этого в начале работы протокола спросим у каждого участника стоимость отрезка $[0, 1]$, а после каждого запроса разрезания, в котором $i$-й участник отрезал кусок $[x, y]$, будем спрашивать у каждого участника дележа, кроме $i$, стоимость $[x, y]$. Таким образом, мы в любой момент будем знать стоимость с точки зрения любого участника всех отрезков, на которые $[0, 1]$ разбивается точками из запросов разрезания, и никакие другие запросы измерения не потребуются.
\end{note}

Далее будем оценивать только количество запросов разрезания.

	\section{Протокол}

Назовём основной протокол, получающий кусок пирога $R$ и множество участников $N$ и строящий делёж без зависти $R$ между участниками из $N$, Main Protocol.

 	\subsection{Обзор протокола и введение обозначений}
Важную роль в алгоритмах дележа без зависти играет понятие доминирования.

\begin{mydef}
Будем говорить, что при частичном дележе $S = (S_1, ..., S_n)$ участник $i$ доминирует над участником $j$, если $v_i(S_i) \geq v_i(S_j) + v_i(R)$, то есть $i$ не будет завидовать $j$, даже если $j$ получит весь остаток.
\end{mydef}

Заметим, что если $S$ "--- частичный делёж без зависти с остатком $R$, при котором участник $i$ доминирует над участником $j$, а $S'$ "--- какой-то частичный делёж $R$ (не обязательно без зависти), то при дележе $S \cup S'$ участник $i$ не будет завидовать участнику $j$ и $i$ будет доминировать над $j$. Таким образом, однажды возникшее доминирование уже не может быть утрачено в ходе протокола.

Как и большинство алгоритмов дележа без зависти, наш протокол будет строить частичные дележи без зависти, добиваясь того, чтобы некоторые участники доминировали над другими участниками, а потом использовать эти отношения для исключения некоторых участников из дележа. Когда множество участников $N$ разбивается на 2 множества $A_1 \sqcup A_2 = N$ такие, что $A_1 \neq \varnothing, A_2 \neq \varnothing$ и при текущем дележе $S$ с остатком $R$ каждый участник из $A_1$ доминирует над каждым участником $A_2$, можно разделить $R$ без завести только между участниками  из $A_2$ с помощью рекурсивного запуска протокола для меньшего числа участников. (Будем называть это "исключить участников из $A_1$ из дележа"). Тогда если $S'$ "--- получившийся делёж $R$ без зависти между участниками из $A_2$, то $S \cup S'$ - делёж без зависти всего пирога между всеми участниками. Для увеличения количества отношений доминирования протокол будет использовать обмены участников кусками.

Если $n \leq 4$, применим один из ранее известных ограниченных протоколов дележа без зависти, далее будем считать, что $n \geq 5$.

Также введём понятие бонуса.

\begin{mydef}
Бонус участника i над участником j при частичном дележе без зависти $S = (S_1, ..., S_n)$ с остатком $R$ "--- величина $v_i(S_i) - v_i(S_j)$, то есть величина, на которую по мнению участника i он получил больше, чем участник j.
\end{mydef}

Заметим, что при дележе без зависти бонус всегда неотрицателен и если бонус участника $i$ над участником $j$ при дележе с остатком $R$ не меньше $v_i(R)$, то участник $i$ доминирует над участником $j$.

Частичные дележи будут строиться с помощью протокола Core protocol, впервые описанного в \cite{AzizMakkenzi}. Этот протокол строит частичный делёж без зависти такой, что один фиксированный участник (будем называть его cutter) получает кусок, равный по его мнению $\frac 1n$ делимого куска, и один из других участников получает кусок, равный по мнению cutter $\frac 1n$ делимого куска. Таким образом, после запуска Core Protocol остаток уменьшается по мнению cutter хотя бы в $\frac {n}{n - 2}$ раза. Если бонус участника $i$ над участником $j$ не меньше, чем $ (\frac {n - 2}{n})^k*v_i(R)$, то его можно превратить в доминирование, запустив $k$ раз Core Protocol(cutter=$i$, текущий остаток). Таким образом, любой ненулевой бонус может быть превращен в доминирование за конечное число запусков Core Protocol. К сожалению, в общем случае это число запусков не ограниченно функцией от $n$. Введём некоторые зависящие только от $n$ константы, с помощью которой мы определим, насколько большой бонус мы будем превращать в доминирование.

Введём обозначения, которые понадобятся нам для дальнейших определений. 
$$ C = n^{4}2^{n^2}n^{3n^2}, C' = C \cdot (n!)^n, B = \log_{\frac {n}{n - 2} } (nC').$$
Заметим, что $C = n^{3n^2(1 + o(1))}$, $C' \leq C \cdot n^{n^2} = n^{4n^2 + o(n^2)}$, $B \leq n\ln (nC') = O(n^4).$

\begin{mydef}
Будем говорить, что при частичном дележе $S = (S_1, ..., S_n)$ и текущем остатке $R$ участник $i$ имеет значительное преимущество над участником $j$, если $v_i(S_i) \geq v_i(S_j) + \frac {1}{nC'} \cdot v_i(R)$, где $R$  "--- текущий остаток.
\end{mydef}

Таким образом, если $i$ имеет значительное преимущество над $j$, то через $B$ запусков Core(cutter=$i$) $i$ будет доминировать над $j$. Поэтому, если мы сможем добиться того, что множество участников $N$ разбивается на 2 множества $A_1 \sqcup A_2 = N$ такие, что $A_1 \neq \varnothing, A_2 \neq \varnothing$ и каждый участник из $A_1$ имеет значительное преимущество над каждым участником из $A_2$, то мы можем за $O(n^5)$ запусков Core Protocol добиться того, что каждый участник из $A_1$ доминирует над каждым участником из $A_2$, для каждого участника $i \in A_1$ запустив $B$ раз Core(cutter=$i$). После этого можно исключить из дележа участников из $A_1$ и завершить делёж рекурсивным запуском Main Protocol для меньшего числа участников.

\begin{mydef}
Если при запуске Core Protocol мы получили частичный делёж $S = (S_1, ..., S_n)$ с остатком $R$, будем называть $(S_1, ..., S_n)$ снимком этого частичного дележа.
\end{mydef}

Заметим, что мы включаем в снимок только куски, полученные участниками при запуске этого Core Protocol (не включаем куски, полученные до или после этого запуска) и не включаем в снимок остаток. В ходе работы протокола остаток будет всё время уменьшаться, но снимки полученных с помощью Core Protocol частичных дележей не будут меняться. При этом то, имеет ли участник $i$ значительное преимущество над $j$ при частичном дележе $(S_1, ..., S_n)$ с остатком $R$, зависит не только от $S_i$ и $S_j$, но и от $R$, при уменьшении $R$ могут появляться новые значительные преимущества.

Также определим понятие значительная величина и значительный кусок.

\begin{mydef}
Будем говорить, что $x \in [0, 1]$ "--- значительная величина для участника  $i$, если $x \geq \frac {1}{nC'} \cdot v_i(R)$, где $R$ "--- текущий остаток.
\end{mydef}

То есть значительная величина "--- любая такая величина $x$, что, если бонус $i$ над каким-то участником равен $x$, то $i$ имеет значительное преимущество над этим участником.

\begin{mydef}
Будем говорить, что кусок $P$ "--- значительный кусок для участника $i$, если $v_i(P)$ является значительной величиной для участника $i$. 
\end{mydef}

Таким  образом, мы умеем работать с являющимися значительной величиной бонусами. В  \cite{AzizMakkenzi} авторы показывают, как с помощью новой техники, названной ими ``извлечение'', работать с бонусами, значительно меньшими, чем минимальное значение значительной величины. Как и в \cite{AzizMakkenzi}, после некоторой подготовки мы будем запускать описанный в \cite{AzizMakkenzi} GoLeft Protocol, который получает новые значительные преимущества для достижения того, что участники из некоторого множества имеют значительное преимущество над остальными участниками. В ходе работы GoLeft Protocol некоторые бонусы могут уменьшиться (однако не могут стать меньше 0, то есть не может появиться зависть). Чтобы при этом уменьшении значительные преимущества не исчезли, нам надо заранее добиться того, что бонусы, которые будут превращены в значительные преимущества, будут значительно больше, чем минимальное значение значительной величины.

\begin{mydef}
Будем говорить, что $x$ "--- незначительная величина для участника i при частичном дележе $S = (S_1, ..., S_n)$ и текущем остатке $R$, если $x \leq \frac {1}{n^2C'} \cdot v_i(R)$, участник $i$ имеет незначительное преимущество над участником $j$ при дележе $S$, если $v_i(S_i) \leq v_i(S_j) + \frac {1}{n^2C'} \cdot v_i(R)$. В частности, если $v_i(S_i) = v_i(S_j)$, то $i$ имеет незначительное преимущество над $j$  и 0 "--- незначительная величина для любого участника при любом частичном дележе без зависти.
\end{mydef}

\begin{mydef} Будем говорить, что $x$ "--- очень значительная величина для участника i при частичном дележе $S = (S_1, ..., S_n)$ и текущем остатке $R$, если $x \geq \frac {1}{C'} \cdot v_i(R)$, участник $i$ имеет очень значительное преимущество над участником $j$ при дележе $S$, если $v_i(S_i) \geq v_i(S_j) + \frac {1}{C'} \cdot v_i(R)$.
\end{mydef}

Неформально, незначительная величина "--- бонус, с которым можно работать с помощью техники извлечения, а очень значительная величина "--- бонус, которое останется значительной величиной при работе GoLeft Protocol.

\includegraphics{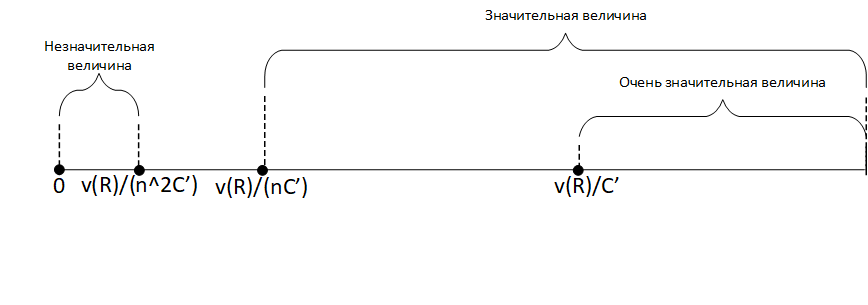}

Наш протокол состоит из двух частей. В первой части, которую мы будем называть PrepareGoLeft Protocol, мы $C'$ раз запускаем Core Protocol (каждый раз делим остаток, полученный предыдущим запуском Core Protocol) и получаем $C'$ снимков. Уменьшая остаток с помощью Core Protocol, мы добиваемся того, что в каждом снимке для каждой пары участников $i$ и $j$ участник $i$ имеет либо очень значительное преимущество над $j$, либо незначительное преимущество над $j$ в этом снимке. Затем мы проводим процедуру извлечения для всех незначительных преимуществ в этих снимках. При этом мы поддерживаем то, что все бонусы в полученных в начале этой фазы $C'$ снимках являются очень значительными величинами или незначительными величинами: если при уменьшении остатка какой-то бонус перестаёт быть незначительной величиной, мы снова уменьшаем остаток с помощью запуска Core Protocol и добиваемся того, чтобы этот бонус стал очень значительной величиной. Иногда в ходе извлечения мы можем сразу разбить задачу на два рекурсивных дележа без зависти для меньшего количества участников. Если этого не происходит, то мы получаем $C'$ снимков, в которых все бонусы являются очень значительными величинами или незначительными величинами и для незначительных величин проведена процедура извлечения. Эти снимки разбиваются на не более чем $(n!)^n$ классов эквивалентности (определение эквивалентных снимков будет дано позже), по принципу Дирихле из них можно выбрать $C$ эквивалентных снимков.

Второй частью протокола является GoLeft протокол. Он принимает $C'$ изоморфных снимков, в которых все бонусы являются очень значительными величинами или незначительными величинами и для незначительных величин проведена процедура извлечения, и, с помощью обменов кусками между участниками и присоединения к кускам из снимков ассоциированных с ними в ходе процедуры извлечения кусков, создаёт новые значительные преимущества и добивается того, что участники из некоторого множества $A_1$ имеют значительное преимущество над остальными участниками. При этом некоторые очень значительные величины могут уменьшиться и перестать быть очень значительными величинами, но останутся значительными величинами. После этого мы, запуская $B$ раз Core Protocol с каждым участником из $A_1$ в качестве cutter, превращаем значительное преимущество в доминирование, исключаем участников из $A_1$ из дележа и рекурсивно запускаем MainProtocol для меньшего количества участников.

 	\subsection{Core protocol}

Core Protocol принимает множество участников $N$, неподелённый кусок пирога $R$ и фиксированного участника $cutter \in N$ и строит такой частичный делёж без зависти $S = (S_1, ..., S_{|N|})$ с остатком $R'$, что $cutter$ и ещё хотя бы один участник получают по мнению $cutter$ ровно $\frac {1}{|N|}$ от $R$. Мы будем использовать без изменения протокол CoreProtocol из \cite{AzizMakkenzi}. Мы не будем заново доказывать его корректность (она доказывается в лемме 4.5 в \cite{AzizMakkenzi}), но улучшим оценку используемого количества запросов с $O(n^{2n+3})$ до $O((\frac{3 + \sqrt{5}}{2})^{n})$. Для того, чтобы показать, что наш MainProtocol работает за экспоненциально время нам было бы достаточно и оценок из \cite{AzizMakkenzi}, но CoreProtocol интересен и сам по себе, например, в \cite{AzizMakkenzi} показано, как с помощью $n$ запусков CoreProtocol построить частичный пропорциональный делёж без зависти и здесь наше улучшение оценки его времени работы уже будет важным.

\subsubsection{Время работы CoreProtocol}

Для удобства приведём реализацию Core Protocol из \cite{AzizMakkenzi}

\includegraphics{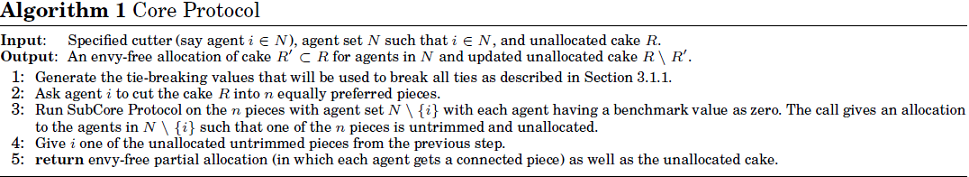}

\includegraphics{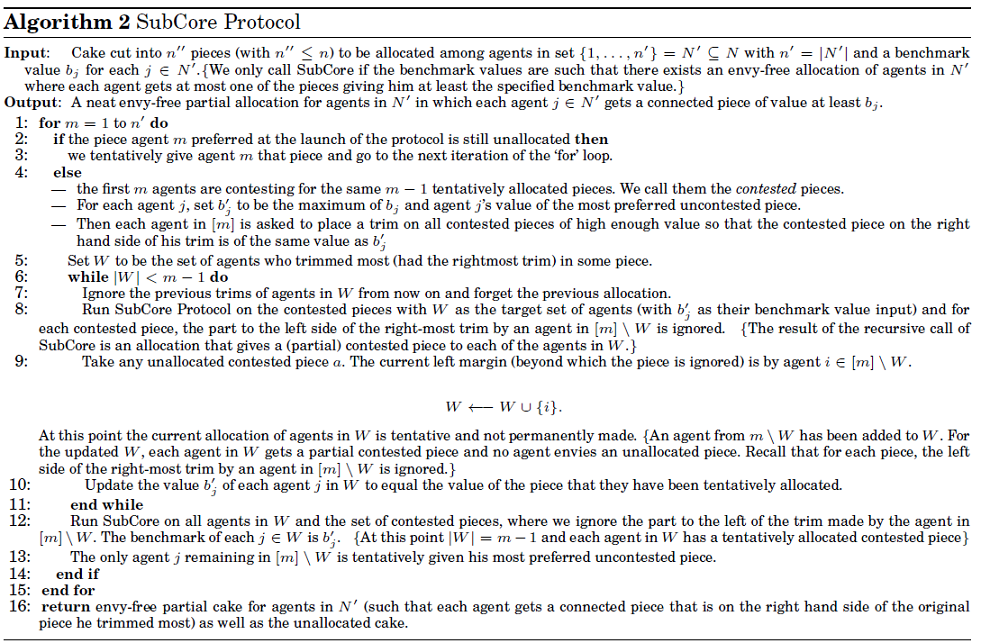}

\begin{lemma}
Core Protocol делает $O((\frac{3 + \sqrt{5}}{2})^{n})$ запросов разрезания и $O(n\cdot(\frac{3 + \sqrt{5}}{2})^{n})$ запросов измерения.
\end{lemma}

\textbf{Доказательство леммы 1}

Core Protocol делает $n - 1$ запрос разрезания, после чего один раз запускает SubCore Protocol с $n - 1$ участниками, поэтому нам достаточно оценить время работы SubCore Protocol.

Пусть $T(n')$ "--- максимальное возможное количество операций разрезания в SubCore Protocol, который делит куски пирога между $n'$  участниками (при оценке $T(n')$ увидим, что наша оценка не зависит ни от каких других параметров SubCore Protocol).

SubCore Protocol добавляет участников по одному в цикле в строке 1. Посчитаем количество операций для каждого $m$ из цикла for в строке 1. В строке 4 для каждого из $m$ уже добавленных участников и каждого из $m - 1$ розданных кусков может быть сделано по 1 запросу разрезания, всего $m(m-1)$ запросов. В цикле while в строке 6 и в строке 12 для каждого $|W|$ от какого-то начального $|W| \geq 1$ до $|W| = m - 1$ рекурсивно запускается SubCore протокол с $|W|$ участниками. Больше в SubCore Protocol не делается запросов разрезания и не происходит рекурсивных запусков. Таким образом,
$$ T(n') \leq \sum_{m = 1}^{n'} (m(m - 1) + \sum_{i = 1}^{m - 1} T(i)).$$
Пусть $$T'(n') = \sum_{m = 1}^{n'} (m(m - 1) + \sum_{i = 1}^{m - 1} T'(i)), \, T'(1) = T(1)$$
Легко показать с помощью индукции по $n'$, что $T(n) \leq T'(n)$, поэтому достаточно оценить $T'(n)$.
$$ T'(n') \leq \sum_{m = 1}^{n'} (m(m - 1) + \sum_{i = 1}^{m - 1} T'(i)) = \sum_{m = 1}^{n' - 1} (m(m - 1) + \sum_{i = 1}^{m - 1} T'(i)) + (n'(n' - 1) + \sum_{i = 1}^{n' - 1} T'(i)) = $$ $$ = T'(n' - 1) + n'(n' - 1) + \sum_{i = 1}^{n' - 1} T'(i) = 2T'(n' - 1) + n'(n' - 1) + \sum_{i = 1}^{n' - 2} T'(i),$$
используя это равенство для $n'$ и $n' - 1$, получаем 
$$ T'(n') = 2T'(n' - 1) + n'(n' - 1) + \sum_{i = 1}^{n' - 2} T'(i) = 2T'(n' - 1) + n'(n' - 1) - T'(n' - 2) - (n' - 2)(n' - 1) + $$ $$ + 2T'(n' - 2) + (n' - 2)(n' - 1) + \sum_{i = 1}^{n' - 3} T'(i) = 3T'(n' - 1) - T'(n' - 2) + 2(n' - 1).$$
Сделав замену переменной $b_{n'} = T'(n') + 2n'$, получаем $b_{n'} = 3b_{n'-1} - b_{n' - 2} \Rightarrow b_{n'} = C_1 (\frac{3 + \sqrt{5}}{2})^{n'} + C_2 (\frac{3 -  \sqrt{5}}{2})^{n'} = O((\frac{3 + \sqrt{5}}{2})^{n'})$, $T'(n') = b_{n'} - 2n' = O((\frac{3 + \sqrt{5}}{2})^{n'})$. $T(n) \leq T'(n)$, поэтому $T(n') = b_{n'} - 2n' = O((\frac{3 + \sqrt{5}}{2})^{n'}).$

Количество запросов разрезания в CoreProtocol $ = n - 1 + T(n - 1) = O((\frac{3 + \sqrt{5}}{2})^{n})$. Из замечания 1 получаем, что можно реализовать CoreProtocol так, что количество запросов измерения в нём не превосходит $ (n - 1) \cdot O((\frac{3 + \sqrt{5}}{2})^{n}) + n = O(n(\frac{3 + \sqrt{5}}{2})^{n})$.

\subsubsection{Свойства CoreProtocol}

Докажем следующие свойства Core Protocol, которые потребуются нам дальше:

\begin{statement}
	При запуске Core Protocol для дележа текущего остатка остаток уменьшается хотя бы в $\frac {n} {n - 2}$ раз по мнению cutter.
\end{statement}

\textbf{Доказательство утверждения 1} Пусть Core Protocol запускается для дележа куска $R$ и возвращает делёж $S = (S_1, ..., S_n)$ с остатком $R'$. Тогда $v_{cutter}(R') = v_{cutter}(R) - \sum_{i = 1}^{n} v_{cutter}(S_i)$, по мнению cutter он и ещё хотя бы один участник дележа получают ровно $\frac 1n$ от $R$, то есть $\exists i \neq cutter : v_{cutter}(S_i) = v_{cutter}(S_{cutter}) = \frac 1n v_{cutter}(R), \Rightarrow \sum_{i = 1}^{n} v_{cutter}(S_i) \geq \frac 2n v_{cutter}(R) \Rightarrow v_{cutter}(R') \leq \frac {n-2}{n} v_{cutter}(R)$.

\begin{statement}
	Если бонус $i$ над $j$ при дележе $S$ не является незначительной величиной, то если $n \ln n$ раз запустить Core Protocol для дележа текущего остатка с cutter=i, то бонус $i$ над $j$ при дележе $S$ станет очень значительной величиной.
\end{statement}

\textbf{Доказательство утверждения 2} Пусть $R$ "--- текущий остаток, $R'$ "--- остаток, который останется после $n \ln n$ запусков Core Protocol для дележа текущего остатка с cutter=i. Если бонус $i$ над $j$ при дележе $S = (S_1, ..., S_n)$ и текущем остатке $R$ не является незначительной величиной, то $v_i(R) \leq n^2C' \cdot (v_i(S_i) - v_i(S_j))$. Из утверждения 1 получаем, что $v_i(R') \leq (\frac {n - 2}{n})^{n \ln n} v_i(R) \Rightarrow \frac {v_i(R)}{v_i(R')} \geq ((1 + \frac {2}{n - 2})^n)^{\ln n} \geq (e^2)^{\ln n} = n^2 \Rightarrow v_i(R') \leq \frac {1}{n^2} v_i(R) \leq C' \cdot (v_i(S_i) - v_i(S_j)) \Rightarrow (v_i(S_i) - v_i(S_j)) \geq \frac {1}{C'}v_i(R')$, то есть бонус $i$ над $j$ при дележе $S$ и текущем остатке $R'$ является очень значительной величиной.

\begin{statement}
	 Если $S$ "--- частичный делёж без зависти, полученный дележом текущего остатка с помощью Core Protocol с cutter=i, то i имеет значительное преимущество при дележе S хотя бы над одним участником.
\end{statement}

\textbf{Доказательство утверждения 3} Пусть $R$ "--- текущий остаток; Core Protocol, запущенный для дележа $R$  с cutter=i, строит делёж $S = (S_1, ..., S_n)$ с остатком $R'$. Тогда $v_i(S_i) = \frac 1n v_i(R) \Rightarrow v_i(R') = v_i(R) - \sum_{j = 1}^{n} v_i(S_j) = n \cdot v_i(S_i) - \sum_{j = 1}^{n} v_i(S_j) = \sum_{j = 1}^{n} (v_i(S_i) - v_i(S_j))$, по принципу Дирихле $\exists j: v_i(S_i) - v_i(S_j) \geq \frac 1n v_i(R')$, для этого $j$ бонус $i$ над $j$ при дележе $S$ не меньше, чем $\frac 1n v_i(R') \geq \frac {1}{nC'} v_i(R')$, то есть $i$ имеет значительное преимущество над $j$.

\begin{statement}
	 Если при дележе $S$ $i$ имеет значительное преимущество над $j$, то после $B$ запусков Core Protocol с cutter=i $i$ будет доминировать над $j$.
\end{statement}

\textbf{Доказательство утверждения 4} Пусть $R$ "--- текущий остаток, $R'$ "--- остаток, который останется после $B$ запусков Core Protocol для дележа текущего остатка с cutter=i. Если при дележе $S$ $i$ имеет значительное преимущество над $j$, то бонус i над j не меньше, чем $\frac{1}{nC'}v_i(R)$, по утвержению 1 $v_i(R') \leq (\frac{n - 2}{n})^B v_i(R) = \frac{1}{nC'}v_i(R)$, поэтому после $B$ запусков Core Protocol с cutter=i бонус $i$ над $j$ будет не меньше, чем стоимость остатка с точки зрения $i$, то есть $i$ будет доминировать над $j$.

 	\subsection{PrepareGoLeft Protocol}

\subsubsection{Обзор PrepareGoLeft Protocol}

PrepareGoLeft Protocol запускается в начале Main Protocol. Он с помощью CoreProtocol генерирует $C'$ снимков, при этом остаётся какой-то остаток. Затем, уменьшая остаток с помощью Core Protocol, он добивается того, что все бонусы являются очень значительными величинами или незначительными величинами. После этого мы запускаем для каждого из $C'$ снимков и каждого участника $j$ процедуру извлечения. В ходе этой процедуры некоторый кусок извлекается участниками, имеющими в этом снимке незначительное преимущество над $j$, из текущего остатка и ассоциируется с куском $j$ в этом снимке. Этот кусок не добавляется к куску $j$ и продолжает считаться частью остатка при определении значительной величины, но больше не является частью остатка при дальнейших извлечениях. В дальнейшем он может быть присоединён к куску $j$ во время GoLeft Protocol или возвращён в остаток во время GoLeft Protocol или при подготовке к процедуре обработки несоответствий. 

Чтобы остатка хватило на извлечение из него кусков во всех $C' \cdot n$ операциях извлечения нам приходится потребовать, чтобы все извлекаемые куски не являлись значительными кусками с точки зрения всех участников. Если какой-то извлекаемый кусок является значительным куском для кого-то из участников, то имеет место значительное несоответствие между стоимостью этого куска по мнению извлекающих его участников и по мнению того, кто считает его стоимость значительной величиной. В этом случае мы возвращаем все остальные извлечённые куски в остаток и запускаем процедуру обработки несоответствий. Эта процедура либо увеличивает число значительных преимуществ, либо использует это несоответствие для разбиения остатка на два куска и множества участников на два непустых множества так, что задача сводится к дележу с помощью двух рекурсивных запусков Main Protocol одного куска между одним множеством участников, а другого куска между другим множеством участников.

Если в результате обработки несоответствий делёж не был завершён, то мы снова добиваемся того, что все бонусы являются очень значительными величинами или незначительными величинами с помощью уменьшения остатка запуском CoreProtocol. (Это является основным отличием нашего протокола от приведённого в \cite{AzizMakkenzi} протокола Азиза-Маккензи собственно в протоколе, а не в его изложении и доказательстве. В оригинальном протоколе Азиза-Маккензи после обработки несоответствий сразу снова запускается извлечение, из-за чего в определении значительной величины, незначительной величины и очень значительной величины (последние два понятия не вводятся в \cite{AzizMakkenzi}, но неявно используются) приходится требовать намного более сильные условия, из-за чего возрастает количество запросов). После этого мы заново запускаем операцию извлечения (напомним, что все ранее извлечённые куски были возвращены в остаток при подготовке к обработке несоответствий). Ограниченность количества запусков извлечения достигается благодаря тому, что после каждого запуска обработки несоответствий появляется новое значительное преимущество в одном из $C'$ сгенерированных в начале PrepareGoLeft Protocol снимков, а их количество не может стать больше, чем $n(n-1)C'$ (количество пар участников, умноженное на количество снимков).

Опишем более подробно процедуры извлечение и обработки несоответствий.

\subsubsection{Извлечение}

В начале PrepareGoLeft Protocol мы генерируем $C'$ снимков, с которыми дальше работаем. Введём обозначение $c_{jk}$ "--- кусок, доставшийся участнику $k$ в $j$-м снимке, $k \in \{1, ..., n\}, j \in \{1, ..., C'\}$. Для каждого снимка и каждого участника мы будем проводить процедуру извлечения.

Заметим, что перед проведением извлечения мы добиваемся того, что все бонусы являются очень значительными величинами или незначительными величинами. При процедуре извлечения остаток не меняется (Core Protocol не запускается, некоторые куски извлекаются из остатка, но при определении незначительной величины и очень значительной величины они продолжают считаться частью остатка). Поэтому во время всех процедур извлечения все бонусы являются очень значительными величинами или незначительными величинами.

При проведении извлечения для снимка $j$ и участника $k$ некоторый кусок извлекается из остатка и ассоциируется с куском $c_{jk}$. Каждый участник, не имеющий значительного преимущества (то есть имеющий незначительное преимущество) над $k$ в снимке $j$ отрезает от правого края остатка кусок, который по его мнению имеет стоимость, равную его бонусу над $k$ в снимке $j$. При этом эти куски пока не извлекаются из остатка и участники игнорируют разрезы друг друга. Обозначим количество участников, не имеющих значительного преимущества над $k$ в снимке $j$, через $m_{jk}$. Таким образом, мы получили на остатке $m_{jk}$ разрезов, стоимость куска справа от разреза, сделанного участником $i$, равна по его мнению его бонусу над $j$. После этого кусок справа от самого левого из сделанных разрезов (на нём сделано $m_{jk} - 1$ разрезов) извлекается из остатка. Тот, кто сделал самый левый разрез, считает, что стоимость этого куска равна его бонусу над $k$ в дележе $j$ и является незначительной величиной, поэтому этот кусок не является по его мнению значительныь куском. Чтобы успешно завершить процедуру извлечения надо, чтобы остальные участники дележа также считали, что этот кусок не является значительным куском.


Если все участники согласны, что этот кусок не является значительным куском, то этот кусок \textbf{ассоциируется} с куском $c_{jk}$. Он кладётся слева от куска $c_{jk}$. Введём несколько обозначений. Ассоциированный кусок разрезан $m_{jk} - 1$ разрезами на $m_{jk}$ кусков. Пронумеруем эти куски справа налево и обозначим $l$-й из них через $e_{jkl}, l \in \{1, ..., m_{jk}\}$. Обозначим через $i_{jkl}$ участника, сделавшего разрез, ограничивающий слева $e_{jkl}$. Будем говорить, что кусок $e_{jkl}$ извлечён участником $i_{jkl}$. Таким образом, справа от разреза, сделанного участником $i_{jkl}$, лежат куски $e_{jk1}, e_{jk2}, ... e_{jkl}$ и сам кусок $c_{jk}$, а слева "--- куски $e_{jk(l+1)}, e_{jk(l+2)}, ... e_{jkm_{jk}}$ и этот участник считает, что суммарная стоимость кусков $e_{jk1}, e_{jk2}, ... e_{jkl}$ равна его бонусу над $k$ в снимке $j$. Эти куски пока не добавляются к куску $c_{jk}$, некоторые из них потом будут присоединены к $c_{jk}$, но большинство будут возвращены в остаток или разделены между частью участников рекурсивным запуском MainProtocol. В ходе GoLeft Protocol мы будем по одному присоединять куски $e_{jkl}$ к $c_{jk}$ справа налево, то есть граница $c_{jk}$ будет двигаться влево (из-за этого GoLeft Protocol так называется). Если к куску $c_{jk}$ присоединён кусок $e_{jkl}$ и все куски правее его, то по мнению $i_{jkl}$ кусок $c_{jk}$ стал равен куску, полученному им в снимке $j$, поэтому он готов обменять свой полученный в снимке $j$ кусок на $c_{jk}$. С помощью таких обменов мы получаем новые значительные преимущества.

После процедуры извлечения будем считать частью снимка $j$ ассоциированные с его кусками $c_{jk}$ куски $e_{jkl}$. Введём понятие изоморфных снимков.

\begin{mydef}
Два снимка $j$ и $j'$ изоморфны, если для любого участника $k$ множества участников, имеющих значительное преимущество над $k$ в снимках $j$ и $j'$ совпадают и порядок, в котором эти участники извлекали куски, совпадает, то есть $\forall k \in \{1, ..., n\} m_{jk} = m_{j'k}$ и $\forall k \in \{1, ..., n\} \forall l \in \{1, ..., m_{jk}\} i_{jkl} = i_{j'kl}$.
\end{mydef}

Если какой-то участник $i$ считает, что извлечённый кусок является значительным куском, то все остальные извлечённые во всех процедурах извлечения куски возвращаются в остаток, все сделанные на всех извлечённых кусках разрезы забываются. Пусть $b$ "--- участник, сделавший разрез, отделяющий извлекаемый сейчас кусок от остального остатка, $A$ "--- извлекаемый сейчас кусок, $R$ "--- весь остальной остаток (в который вернули все извлечённые куски, кроме $A$). Тогда $i$ считает, что $A$ является значительным куском, то есть его стоимость не меньше, чем $\frac {1}{nC'}$ от стоимости $A \cup R$, а $b$ считает, что стоимость $A$ является незначительной величиной, то есть не превосходит $\frac {1}{n^2C'}$ от стоимости $A \cup R$. Таким образом имеется сильное разногласие между $i$ и $b$ относительно ценности куска $A$. Для того чтобы использовать это разногласие, мы запускаем процедуру обработки несоответствий, которая либо увеличивает число значительных преимуществ (а именно приводит к тому, что $b$ получает значительное преимущество над $k$ в дележе $j$, хотя раньше не имел его, так как считал свой бонус над $k$ в этом дележе, равный стоимости $A$ по построению $A$, незначительной величиной), либо сводит задачу к двум рекурсивным запускам Main Protocol для меньшего числа участников, для дележа $A$ между группой участников, в которую входит $i$, и дележа части $R$ между группой участников, в которую входит $b$.

\subsubsection{Обработка несоответствий}

Во время процедуры извлечения может получиться так, что кто-то считает извлекаемый кусок значительным куском. В этом случае все остальные извлечённые куски возвращаются в остаток и запускается процедура обработки несоответствий. Пусть $A$ "--- извлекаемый кусок, $c_{jk}$ "--- кусок, к которому мы хотели его присоединить, $b$ "--- участник, сделавший разрез, ограничивающий слева этот кусок (он считает, что стоимость этого куска является незначительной величиной), $i$ "--- участник, считающий этот кусок значительным куском, $R$ "--- остальной остаток. Мы запускаем $B$ раз Core Protocol с cutter=i для дележа $R$. После этого $R$ уменьшается с точки зрения $i$ хотя бы в $nC'$ раз $\Rightarrow$ теперь $i$ считает, что $A$ не меньше, чем $R$. После этого, продолжая уменьшать $R$ с помощью запуска Core Protocol, мы добиваемся того, что все участники считают, что $A$ хотя бы в $n$ раз больше, чем $R$ или хотя бы в $n$ раз меньше, чем $R$.

Если $b$ считает, что $A$ больше, чем $R$, то теперь $b$ имеет значительное преимущество над $k$ в дележе $j$. Тогда мы возвращаем $A$ в остаток и запускаем с начала процедуру извлечения. В ходе обработки несоответствий мы уменьшали остаток с помощью Core Protocol, поэтому часть незначительных преимуществ могла перестать быть незначительными преимуществами, поэтому перед запуском процедуры извлечения мы снова добиваемся того, что все бонусы являются очень значительными величинами или незначительными величинами.

Если $b$ считает, что $R$ больше, чем $A$, то мы разбиваем множество участников $N$ на две части: $N_1$ "--- множество участников, считающих что $A$ больше, чем $R$, хотя бы в $n$ раз и $N_2$ "--- множество участников, считающих что $R$ больше, чем $A$ хотя бы в $n$ раз. Мы добились того, что каждый участник считает, что один из кусков $A$ и $R$ хотя бы в $n$ раз больше, чем другой, то есть $N = N_1 \sqcup N_2$. При этом $N_1 \neq \varnothing$ и $N_2 \neq \varnothing$, так как $i \in N_1, b \in N_2$. После этого мы делим без зависти кусок $A$ между участниками из $N_1$, а кусок $R$ между участниками из $N_2$ с помощью двух рекурсивных запусков Main Protocol для меньшего числа участников и завершаем делёж.

\subsubsection{Реализация PrepareGoLeft Protocol}

PrepareGoLeft Protocol запускается в начале Main Protocol. Он принимает множество участников $N$ и кусок пирога $R$ и возвращает булеву переменную $a$, показывающую, был ли делёж полностью завершён в ходе PrepareGoLeft Protocol. Если $a=1$, он делит без зависти кусок пирога $R$ без остатка. Если $a=0$, он также возвращает множество из $C$ изоморфных снимков, в которых проведена процедура извлечения, остаток $R$ и делит без зависти остальную части пирога.

После запуска Core Protocol для дележа текущего остатка $R$ переобозначаем $R = $ остаток, который вернул этот запуск Core Protocol. Везде, где мы говорим выполнить какую-то операцию нецелое число раз (например, $n \ln n$ раз запустить Core Protocol) подразумевается округление этого числа раз в большую сторону.

Мы считаем, что в Main Protocol определён глобальный частичный делёж без зависти $S_R$. Этот делёж состоит из уже розданных участникам кусков, с которыми больше ничего не будет происходить, в конце Main Protocol этот делёж должен стать дележом без зависти и без остатка, который Main Protocol вернёт. В начале PrepareGoLeft Protocol этот делёж пуст. Когда в строках 2-34 мы строим какие-то дележи без зависти (запускаем Core Protocol или Main Protocol для меньшего числа участников), эти дележи добавляются в $S_R$.

1. $C'$ раз запустить Core Protocol($R$), запомнить сгенерированные $C'$ снимки. Пусть $p_j$ "--- полученный $j$-м запуском Core Protocol снимок, $c_{jk}$ "--- кусок, который $k$-й участник получил в этом снимке. В дальнейшем Core Protocol запускается только для уменьшение остатка, полученные последующими запусками снимки не добавляются в множество снимков. \\
2. Line2:  (в это место мы будем возвращаться, если процедура обработки несоответствий не привела к двум рекурсивным запускам Main Protocol) \\
3. While ($\exists j, i, k: \frac{1}{n^2C'}v_i(R) \leq v_i(c_{ji}) - v_i(c_{jk}) \leq \frac{1}{C'}v_i(R)$) do \\
4. $n \ln n$ раз запустить Core Protocol($R$, cutter=i) \\
(Пока бонус какого-то участника $i$ над каким-то участником $k$ не является очень значительной величиной или незначительной величиной, запускаем Core Protocol с cutter=i для уменьшения остатка) \\
5. end while(3) \\
(Начало извлечения) \\
6. for $(j, k) \in \{1, ..., C'\} \times \{1, ..., n\}$ do \\
7. for $i \in \{1, ..., n\} \setminus {k}$ do \\
8. if $i$ имеет незначительное преимущество над $k$ в снимке $p_j$ then \\
9. $i$ отрезает справа от $R$ кусок, стоимость которого равна по мнению $i$ его бонусу на $k$ в снимке $p_j$. Этот кусок пока не отделяется от $R$, другие участники в цикле на строках 7-11 игнорируют этот разрез. \\
10. end if (8) \\
11. end for (7) \\
12. Пусть $m_{jk}$ "--- количество участников, для которых выполнилось условие в строке 8. Пусть $A$ "--- кусок справа от самого левого из сделанных в строке 9 разрезов. $A$ разбивается другими сделанными в строке 9 разрезами на $m_{jk}$ кусков. Пронумеруем эти куски справа налево и обозначим $l$-й из них через $e_{jkl}$. Кусок $A$ извлекается из остатка $R$, кладётся слева от $c_{jk}$ и ассоциируется с $c_{jk}$. \\
13. if какой-то участник $i$ считает, что кусок $A$ является значительным куском \\
(Начало обработки несоответствий) \\
14. Все ранее извлечённые куски (но не извлекаемый сейчас кусок $A$) возвращаются в остаток. Теперь весь неподеленный пирог разделен на куски $A$ и $R$. \\
15. Пусть $i$ "--- участник, считающий, что $A$ является значительным куском, $b$ "--- участник, который отрезал кусок $A$ в строке 9. \\
16. $B$ раз запустить Core($R$, cutter=i) \\
17. while ($\exists r \in \{1, ..., n\}: \frac 1n v_r(A) \leq v_r(R) \leq nv_r(A)$) do \\ 
18. $n \ln n$ раз запустить Core($R$, cutter=r) \\
19. end while (17) \\
20. if $v_b(R) \geq v_b(A) $ then \\
21. Пусть $N_1$ "--- множество участников, считающих, что $A$ больше, чем $R$, $N_2$ "--- множество участников, считающих, что $R$ больше, чем $A$. \\
22. Запустить Main($A$, $N_1$) \\
23. Запустить Main($R$, $N_2$) \\
24. $a := 1$ \\
25. return $a$; \\
26. else (20) \\
27. $R := R \cup A$ (возвращаем $A$ в остаток $R$) \\
28. GOTO Line2 \\
29. end if (20) \\
(Конец обработки несоответствий) \\
30. end if (13) \\ 
31. end for (6) \\
(Конец извлечения) \\
32. Выбрать из созданного в строке 1 множества из $C'$ снимков $C$ изоморфных снимков. Куски, ассоциированные с кусками из всех остальных $C' - C$ снимков, возвращаются в остаток. \\
33. $a := 0$ \\
34. return $a$, $R$, выбранное в строке 32 множество из $C$ изоморфных снимков. \\ 

\subsubsection{Доказательство корректности PrepareGoLeft Protocol}

Покажем, что все операции в PrepareGoLeft Protocol выполнимы и если он завершается за конечное число запросов, то если он возвращает $a = 1$, то он строит делёж $R$ без зависти, а если он возвращает $a = 0$, то он строит частичный делёж $R$ без зависти и возвращает множество из $C$ изоморфных снимков, в которых проведена процедура извлечения. Оценка количества операций в GoLeftProtocol (в том числе доказательство его конечности) будет приведено в следующем разделе.

Мы будем доказывать корректность Main Protocol индукцией по $n$ (база индукции для $n \leq 3$ следует из корректности используемых при $n \leq 3$ ранее известных протоколов). Поэтому при доказательстве корректности мы будем предполагать, что MainProtocol для любого меньшего числа участников строит делёж без зависти, во всех леммах из этого раздела это предполагается по умолчанию.

\begin{lemma}
В строке 9 кусок $R$ по мнению $i$ стоит не меньше, чем бонус $i$ над $k$ в снимке $p_j$. (То есть кусок $R$ достаточно большой, чтобы $i$ смог отрезать от него кусок, стоимость которого равна по его мнению его бонусу над $k$ в снимке $p_j$).
\end{lemma}

\textbf{Доказательство леммы 2} Всего от $R$ отрезают $nC'$ кусков. Если какой-то из этих кусков является по мнению $i$ значительным куском, то срабатывает условие в строке 13 и процедура извлечения завершается. Если все эти куски не являются по мнению $i$ значительным куском, то стоимость каждого из них меньше, чем $\frac {1}{nC'} v_i(R)$, поэтому их суммарная стоимость по мнению $i$ меньше, чем $nC' \cdot \frac {1}{nC'} v_i(R) = v_i(R) \Rightarrow$ оставшаяся часть куска $R$ всегда будет достаточно большой, чтобы отрезать от неё очередной такой кусок.

\begin{lemma}
Получившийся после выполнения строки 23 делёж является дележом без зависти.
\end{lemma}

\textbf{Доказательство леммы 3} Перед рекурсивными запусками Main Protocol в строках 22 и 23 никто никому не завидовал, потому что все получали куски пирога только в ходе запусков Core Protocol, который проводил частичные дележи без зависти. В строках 22 и 23 мы запускаем Main Protocol с меньшим числом участников, который по предположению индукции строит корректные дележи без зависти между участниками из $N_1$ и между участниками из $N_2$, поэтому участники из $N_1$ не завидуют друг другу и участники из $N_2$ не завидуют друг другу. Из отсутствия зависти в дележе без остатка следует пропорциональность, поэтому каждый участник из $N_1$ получает в строке 22 хотя бы $\frac {1}{|N_1|} \geq \frac 1n$ от куска $A$. По построению $N_1$ он считает, что $A$ больше, чем $R$, из-за того, что цикл while в строках 17-19 завершился, из этого следует, что он считает, что стоимость $A$ не менее чем в $n$ раз больше стоимости $R$, поэтому стоимость полученного им в строке 22 куска не меньше, чем стоимость всего куска $R$ и тем более не меньше, чем стоимость любой части $R$ полученной кем-то из $N_2$ в строке 23. Поэтому никто из $N_1$ не завидует никому из $N_2$. Аналогично каждый участник из $N_2$ не завидует никому из $N_1$, потому что считает, что доставшийся ему в строке 23 кусок не меньше, чем весь кусок $A$, поделенный в строке 22 между участниками из $N_1$.

Введённое в параграфе "извлечение" \ отношение изоморфизма на множестве снимков, в которых проведена процедура извлечения, очевидно является отношением эквивалентности и разбивает снимки на классы изоморфности.

\begin{lemma}
Количество классов изоморфности снимков не превосходит $(n!)^n$.
\end{lemma}

\textbf{Доказательство леммы 4} Класс изоморфности, к которому принадлежит снимок, зависит от того, какие участники сделали разрезы на каждом из $n$ кусков этого снимка в строке 9 и от порядка этих разрезов. Для каждого куска есть $n-1$ участник, который может сделать разрез на этом куске. Для фиксированного куска если на нём было сделанно $k$ разрезов, то есть $\frac {(n - 1)!}{(n - 1 - k)!}$ вариантов того, кем они были сделаны и в каком порядке расположенны. Всего для одного куска есть $\sum_{k = 0}^{n - 1} \frac {(n - 1)!}{(n - 1 - k)!} \leq n \cdot (n - 1)! = n!$ вариантов того, кем сделаны и в каком порядке расположены надрезы на этом куске. Всего $n$ кусков, поэтому всего есть не более, чем $(n!)^n$ классов изоморфности.

\begin{lemma}
В строке 32 из $C'$ снимков можно выбрать $C$ изоморфных снимков.
\end{lemma}

\textbf{Доказательство леммы 5} По определению $C' = (n!)^n \cdot C$, по лемме 4 количество классов изоморфности снимков не превосходит $(n!)^n$, поэтому по принципу Дирихле из $C'$ снимков можно выбрать $C$ изоморфных снимков.

Из лемм 2, 3, 5 и леммы 4.5 из \cite{AzizMakkenzi} (о корректности Core Protocol) следует следующая теорема

\begin{theorem}
Если для любого меньшего, чем $n$, числа участников Main Protocol строит делёж без зависти, то PrepareGoLeft Protocol для $n$ участников успешно завершается и если он возвращает $a = 1$, то он строит делёж пирога без зависти между $n$ участниками, а если он возвращает $a=0$, то он также возвращает множество из $C$ изоморфных снимков, в которых проведена процедура извлечения, и строит частичный делёж остального пирога без зависти.
\end{theorem}

\subsubsection{Оценка времени работы PrepareGoLeft Protocol}

Докажем следующую теорему, оценивающую число запросов в PrepareGoLeft Protocol.

\begin{theorem}
Если $T_c$ "--- такая функция, что количество запросов разрезания, которые делает Core Protocol для $k$ участников не превосходит $T_c(k)$, то PrepareGoLeft Protocol для $n$ участников не более, чем 2 раза вызывает Main Protocol для меньшего числа участников и кроме этого делает не более, чем $C' \cdot T_c(n) + n^2C' \cdot n \ln n \cdot T_c(n) + n^2C' \cdot (n^2C' + B \cdot T_c(n) + n^2 \ln n \cdot T_c(n))$ операций разрезания.
\end{theorem}

Из этой теоремы и леммы 1 будет следовать следующая теорема

\textbf{Теорема 2'}  PrepareGoLeft Protocol для $n$ участников не более, чем 2 раза вызывает Main Protocol для меньшего числа участников и кроме этого делает не более, чем $n^{8n^2(1 +o(1))}$ запросов разрезания.

После выполнения строки 1 введём полуинвариант, равный количеству таких троек $(j, i, k) \in \{1, ..., C'\} \times \{1, ..., n\} \times \{1, ..., n\}$, что $i$ имеет очень значительное преимущество над $k$ в снимке $p_j$.

\begin{lemma}
Полуинвариант не уменьшается и может увеличиться не более, чем $n^2C' - 1$ раз.
\end{lemma}

\textbf{Доказательство леммы 6} В ходе PrepareGoLeft Protocol остаток никогда не увеличивается (мы можем возвращать в остаток извлечённые из него куски, но они и так считаются частью остатка при определении очень значительной величины), из определения очень значительной величины следует, что если какая-то величина является для кого-то очень значительной величиной, то при уменьшении остатка она останется очень значительной величиной, поэтому если какой-то участник получил над кем-то очень значительное преимущество в каком-то снимке, то после этого он всегда будет иметь очень значительное преимущество над этим участником в этом снимке, поэтому полуинвариант никогда не уменьшается. Полуинвариант может принимать значания от 1 до $n^2C'$ (он не может быть равен 0 из утверждения 3 о свойствах Core Protocol) и никогда не уменьшается, поэтому он может увеличиться не более, чем $n^2C' - 1$ раз.

\begin{lemma}
В ходе всей работы протокола цикл while в строках 3-5 выполняется не более, чем $n^2C'$ раз.
\end{lemma}

\textbf{Доказательство леммы 7} Из утверждения 2 о свойствах Core Protocol следует, что если условие в строке 3 выполнилось для каких-то $j, i, k$, то $i$ не имеет очень значительного преимущества над $k$ в снимке $p_j$, но получит его после выполнения строки 4, поэтому при каждом выполнении этого цикла полуинвант увеличивается, по лемме 6 он может увеличиться не более, чем $n^2C'$ раз, поэтому этот цикл выполняется не более, чем $n^2C'$ раз.

\begin{lemma}
Оператор GOTO в строке 28 выполняется не более, чем $n^2C' - 1$ раз. (То есть процедура извлечения начинает выполняться с начала не более, чем $n^2C'$ раз).
\end{lemma}

\textbf{Доказательство леммы 8} Если выполнилась строка 28, то не выполнилось условие if в строке 20, то есть $v_b(A) > v_b(R) \Rightarrow v_b(A) > \frac {1}{C'} (v_b(A) + v_b(R))$, то есть $b$ считает, что стоимость $A$ является очень значительной величиной. По построению он считает, что стоимость $A$ равна его бонусу над $k$ в снимке $p_j$, то есть имеет очень значительное преимущество над $k$ в снимке $p_j$. Но когда он отрезал кусок $A$ от остатка он имел незначительное преимущество над $k$ в снимке $j$ (так как выполнилось условие в строке 8), то есть перед каждым выполнением строки 28 полуинвариант увеличивается. Полуинвариант может увеличиваться не более, чем $n^2C' - 1$ раз, поэтому строка 28 выполняется не более, чем $n^2C' - 1$ раз.

\begin{lemma}
Во время каждой процедуры извлечения цикл while в строках 17-19 выполняется не более, чем $n$ раз.
\end{lemma}

\textbf{Доказательство леммы 9} При каждом выполнении строки 18 участник $r$, для которого в строке 17 выполнялось $\frac 1n v_r(A) \leq v_r(R) \leq nv_r(A)$, начинает считать, что стоимость $A$ хотя бы в $n$ раз больше стоимости $R$ (это выводится из утверждения 1 о свойтвах Core Protocol так же, как утверждение 2 о свойствах Core Protocol) и продолжает так считать, так как остаток $R$ только уменьшается, поэтому для каждого участника дележа строка 18 выполняется не более одного раза.

\textbf{Доказательство теоремы 2} В строке 1 делается не более, чем $C' \cdot T_c(n)$ операций разрезания. При каждом запуске цикла while в строках 3-5 выполняется не более чем $n \ln n \cdot T_c(n)$ операций разрезания, по лемме 7 этот цикл выполняется не более, чем $n^2C'$ раз $\Rightarrow$ всего в строках 3-5 выполняется не более, чем $n^2C' \cdot n \ln n \cdot T_c(n)$ операций разрезания. При одном запуске извлечения строка 9 выполняется не более, чем $n^2C'$ раз, строка 18 по лемме 9 выполняется при одном запуске извлечения не более, чем $n$ раз и при каждом её запуске выполняется не более, чем $n \ln n \cdot T_c(n)$ операций разрезания, в строке 16 при каждом запуске извлечения выполняется не более, чем $B \cdot T_c(n)$ операций разрезания, поэтому всего в строках 6-21 при каждом запуске извлечения выполняется не более $n^2C' + B \cdot T_c(n) + n^2 \ln n \cdot T_c(n)$ операций разрезания. По лемме 8 извлечение запускается не более, чем $n^2C'$ раз, поэтому всего в строках 6-21 выполняется не более, чем $n^2C' \cdot (n^2C' + B \cdot T_c(n) + n^2 \ln n \cdot T_c(n))$ операций разрезания. Строки 22 и 23 выполняются не более одного раза (так как после их выполнения делёж завершается), в каждой из них выполняется не более $T_m(n - 1)$ операций разрезания. Других операций разрезания в PrepareGoLeft Protocol нет, поэтому всего в PrepareGoLeft Protocol выполняется не более, чем $C' \cdot T_c(n) + n^2C' \cdot n \ln n \cdot T_c(n) + n^2C' \cdot (n^2C' + B \cdot T_c(n) + n^2 \ln n \cdot T_c(n)) + 2T_m(n - 1)$ операций разрезания.

\textbf{Доказательство теоремы 2'} Подставив в теорему 2 $T_c(n) = O((\frac {\sqrt{5} + 3}{2})^n)$ и $C' = n^{4n^2(1 + o(1))}$ получаем, что кроме двух рекурсивных вызовов Main Protocol для меньшего числа участников PrepareGoLeft Protocol выполняет не более, чем $n^{8n^2(1 + o(1))}$ операций разрезания.

 	\subsection{GoLeft protocol}

\subsubsection{Обзор GoLeft Protocol}

GoLeft Protocol получает на вход $C$ изоморфных снимков, в которых проведена процедура извлечения и в которых все бонусы являются очень значительными величинами или незначительными величинами, и остаток $R$. Также к моменту запуска GoLeft Protocol часть пирога уже разделена без зависти, обозачим этот частичный делёж без зависти через $S_R = (S_{R1}, ..., S_{Rn})$ (он является объединением снимков всех частичных дележей, полученных с помощью Core Protocol в PrepareGoLeft Protocol, кроме тех $C$ изоморфных снимков, которые он вернул). GoLeft Protocol добивается того, что множество участников разбивается на два непустых множества $A_1$ и $A_2$ такие, что каждый участник из $A_1$ имеет значительное преимущество над каждым участником из $A_2$. После этого, запуская $B$ раз Core Protocol с каждым участником в качестве $cutter$, мы превращаем значительные преимущества в доминирования, исключаем из дележа участников из $A_1$ и завершаем делёж рекурсивным запуском Main Protocol для меньшего множества участников $A_2$.

GoLeft Protocol работает с множеством изоморфных активных снимков, которое изначально состоит из $C$ полученных на вход изоморфных снимков. В ходе работы протокола это множество уменьшается. Основными операциями GoLeft Protocol являются присоединения ассоциированных кусков $e_{jkl}$ к кускам $c_{jk}$ из дележа и обмены кусками в активных снимках. При этом все операции происходят одновременно для всех активных снимков: если для какого-то $j$ кусок $e_{jkl}$ присоединяется к куску $c_{jk}$, то одновременно для всех $j$, для которых $j$-й снимок ещё является активным, кусок  $e_{jkl}$ присоединяется к куску $c_{jk}$ и если в каком-то активном снимке участники $k$ и $i$ меняются кусками, то и во всех активных снимках они меняются кусками. Обозначим множество всех находящихся в активных снимках кусков $c_{jk}$ через $c_{...k}$, а множество всех находящихся в активных снимках кусков $e_{jkl}$ через $e_{...kl}$. Таким образом всё, что происходит с куском $c_{jk}$ или $e_{jkl}$ (кроме удаления из множества активных снимков) происходит одновременно и со всеми кусками из $c_{...k}$ или $e_{...kl}$. Напомним, что мы обозначили через $i_{jkl}$ участника, который извлёк кусок $e_{jkl}$. Заметим, что все активные снимки изоморфны, поэтому для любых $j, j', k, l$ $i_{jkl} = i_{j'kl}$, будем обозначать участника, который извлёк все куски из $e_{...kl}$ через $i_{kl}$.

Благодаря обменам появляются новые значительные преимущества: если какой-то участник $i$ имел в активных снимках значительное преимущество над участником $k$, а в результате обмена участник $k$ отдаёт свои куски участнику $k'$, $i$ получает значительное преимущество над $k'$. Чтобы при обменах не появлялась зависть, мы будем присоединять к кускам из $c_{...k}$ ассоциированные с ними куски из $e_{...kl}$. Для каждого куска $c_{jk}$ ассоциированныые с ним куски $e_{jkl}$ будут присоединяться справа налево (то есть в порядке возрастания $l$), таким образом граница между куском $c_{jk}$ и ассоциированными с ним кусками будет двигаться влево. Чтобы при присоединении ассоциированных кусков не возникало зависти и при обменах сохранялись старые значительные преимущества, мы будем удалять некоторые снимки из множества активных снимков и фиксировать получившийся в них делёж. При этом куски, присоединённые к удаляемым из множества активных снимков кускам, будут возвращаться в остаток или делиться между частью участников дележа с помощью рекурсивного запуска Main Protocol с меньшим числом участников. Константа $C$ выбрана достаточно большой, чтобы активные снимки не кончились при этих удалениях.

После присоединения кусков $e_{...kl}$ к кускам $c_{...k}$ этот кусок будет отдаваться в ходе некоторого циклического обмена участнику $i_{kl}$. При этом те участники, которые в начале GoLeft Protocol имели значительное преимущество над $k$ в активных дележах (то есть те, кто не извлекали куски, ассоциированные с $c_{...k}$) получают значительные преимущества над $i_{kl}$. С помощью удаления снимков из множества активных снимков мы добьёмся того, что значительные преимущества не изчезают при присоединениях и обменах, поэтому участники, которые имели значительное преимущество над $k$ будут иметь значительное преимущество над $k$ и над всеми участниками, чьи куски $e_{...kl}$ уже присоединены к $c_{...k}$. Когда для какого-то $k$, для которого есть хотя бы один участник, имевший в начале GoLeftProtocol значительное преимущество над $k$ в активных снимках, все ассоциированные с $c_{...k}$ куски $e_{...kl}$ оказываются присоединены к $c_{...k}$, каждый участник, имевший в начале GoLeft Protocol значительное преимущество над $k$ в активных снимках, будет иметь значительное преимущество над каждым участником, не имевшим в начале GoLeft Protocol значительного преимущества над $k$. Таким образом все участники разобьются на два непустых множества $A_1$ и $A_2$ такие, что каждый участник из $A_1$ имеет значительное преимущество над каждым участником из $A_2$.

\subsubsection{Обозначения и инварианты}

Пусть $A$ "--- множество номеров активных дележей, то есть множество таких j, что снимок $(c_{j1}, ..., c_{jn})$ ещё не удалён из множества активных снимков. При этом $A$ не меняется при присоединении кусков и обменах, если в активном снимке $(c_{j1}, ..., c_{jn})$ произошли присоединения или обмены, мы по-прежнему будем считать, что $(c_{j1}, ..., c_{jn})$ лежит в множестве активных снимков и $j \in A$. Изначально $A = \{1, ..., C\}$ (считаем, что в начале GoLeft Protocol или в  конце PrepareGoLeft Protocol мы перенумеровали $C$ изоморфных снимков, которые вернул GoLeft Protocol, пронумеровали из числами от 1 до $C$).

Ведём величину $b_k := \sum_{j \in A} v_k(c_{jk})$. Здесь под $c_{jk}$ также подразумеваем кусок, который участник $k$ имел в снимке $j$ в начале GoLeft Protocol, $b_k$ не меняются при присоединениях и обменах, меняются только при уменьшении множества активных дележей. $b_k$ "--- стоимость для $k$ тех кусков, которыми он изначально (в начале GoLeft Protocol) владел в тех снимках, которые ещё являются активными. Также обозначим через $b_{jk}$ стоимость для $k$ куска, которым он изначально владел в снимке $k$ (то есть $b_{jk} := v_k(c_{jk})$).

Пусть $S_A = (S_{A1}, ..., S_{An})$ "--- текущий делёж в активных снимках (в начале GoLeft Protocol $S_{Ak} = \bigcup_{j \in A} c_{jk}$, $S_{Ak}$ меняются при присоединениях ассоциированных кусков, обменах и удалении снимков из множества активных снимков).

Чтобы сформулировать поддерживаемые в GoLeft Protocol инварианты введём бонусный делёж $S_B = (S_{B1}, ..., S_{Bn})$, "--- частичный делёж без зависти, в которые мы будем добавлять снимки, удалённые из множества активных снимков, и дележи, полученные рекурсивным запуском Main Protocol для меньшего числа участников. В начале GoLeft Protocol $ \forall i S_{Bi} = \varnothing$. Из-за обменов и присоединений $S_A$ может не являться дележём без зависти, но $S_A \cup S_{B}$ всегда будет являться дележём без зависти. Заметим, что мы не производим никаких операций над кусками из дележа $S_B$. Таким образом этот делёж вводится только для удобства доказательства протокола, при реализации можно не вводить его, а то, что мы добавляем к нему, сразу раздавать участникам дележа (то есть добавлять к $S_R$).

Пусть $bv_{ii'} := v_i(S_{Bi}) - v_i(S_{Bi'})$ "--- бонус $i$ над $i'$ в дележе $S_B$.

Введём поддерживаемые в GoLeft Protocol инварианты.

\textbf{Инвариант 1}
Каждый участник $k$ имеет в каждом активном снимке $p_j$ хотя бы $b_{jk}$, то есть хотя бы столько, сколько имел в начале GoLeft Protocol

\textbf{Следствие из инварианта 1} 
$v_k(S_{Ak}) \geq b_k$, то есть каждый участник имеет в дележе $S_A$ хотя бы столько, сколько имел во входящих в $S_A$ активных снимках в начале GoLeft Protocol.

\textbf{Инвариант 2} 

Пусть в какой-то момент времени к кускам из $c_{...k}$ присоединены $g_k$ кусков из $e_{...k1}, ..., e_{...km_k}$. Пусть $i$ и $i'$ "--- некоторые участники дележа, $l_{k}$ "--- номер кусков, ассоциированных с $c_{...k}$, извлечённых $i$ (то есть $i = i_{kl_{k}}$), если $i$ не извлекал ассоциированные с $c_{...k}$ куски, то $l_{k}$ не определенно, аналогично $l'_{k}$ "--- номер кусков, ассоциированных с $c_{...k}$, извлечённых $i'$, если $i'$ извлекал такие куски. Для удобства считаем, что $i_{k0} = k$, если $i=k$, то $l_k = 0$ (нам не важно, владел ли участник куском изначально или уже присоединил к нему свой ассоциированный кусок). \\
Пусть $M_1$ "--- множество таких k, что $l'_k < g_k < l_k$ (то есть $i$ извлекал ассоциированные с $c_{...k}$ куски $e_{...kl_k}$ и эти куски ещё не были присоединены, а $i'$ извлекал ассоциированные с $c_{...k}$ куски $e_{...kl'_k}$ и эти куски были присоединены, после этого $i'$ владел кусками $c_{...k}$ с присоединёнными к ним кусками $e_{...k1}, ..., e_{...kl'_k}$, но уже отдал их участнику $i_{k(l'_k+1)}$ после присоединения $e_{...k(l'_k+1)}$). \\
Пусть $M_2$ "--- множество таких $k$, что $l_k < g_k$ и $g_k \leq l'_k$ или $i'$ имел в начале GoLeft Protocol значительное преимущество над $k$ (то есть $i'$ не извлекал ассоциированные с $c_{...k}$ куски $e_{...kl_k}$ или извлекал их, но они ещё не были присоединены или $i'$ сейчас владеет кусками $c_{...k}$, а $i$ извлекал ассоциированные с $c_{...k}$ куски $e_{...kl'_k}$ и эти куски уже были присоединены и $i$ уже не владеет $c_{...k}$). \\
Пусть $M_3$ "--- множество таких $k$, что $l'_k < g_k$ и $i$ не извлекал ассоциированные с $c_{...k}$ куски. \\
Тогда $bv_{ii'} \geq \sum_{k \in M_1} \sum_{j \in A} \sum_{t = g_k + 1}^{l_k} v_i(e_{jkt}) +
 \sum_{k \in M_2} \sum_{j \in A} \sum_{t = l_k + 1}^{g_k} v_i(e_{jkt}) + 
\sum_{k \in M_3} \sum_{j \in A} (\sum_{t = g_k + 1}^{m_k} v_i(e_{jkt}) + \frac {1}{nC'}v_i(R))$.

\begin{note}
В GoLeft Protocol остаток может уменьшаться при присоединении ассоциированных кусков и при их дележе в ходе рекурсивных запусков Main Protocol. В описании и доказательстве корректности GoLeft Protocol мы будем подразумевать под $v_i(R)$ стоимость остатка в начале GoLeft Protocol и под значительным преимуществом, незначительным преимуществом и очень значительным преимуществом будем подразумевать значительное преимущество, незначительное преимущество и очень значительное преимущество относительно остатка, бывшего в начале GoLeft Protocol. Это нужно для того, чтобы незначительные преимущества оставались незначительными преимуществами в течении всего GoLeft Protocol. Однако если мы доказали, что какой-то участник имеет значительное преимущество над другим участником относительно остатка, бывшего в начале GoLeft Protocol, то он имеет значительное преимущество над этим участника и относительно текущего остатка, так как остаток не увеличивается, а значительное преимущество сохраняется при уменьшении остатка.
\end{note}

Заметим, что из инварианта 2 следует, что все $bv_{ii'}$ неотрицательны, то есть $S_B$ "--- делёж без зависти.

Позже мы покажем, что эти инварианты выполняются в GoLeft Protocol и что из них следует, что $S_A \cup  S_B$ "--- делёж без зависти и каждый участник, не извлекавший ассоциированные с $c_{...k}$ куски имеет значительное преимущество в этом дележе над каждым участником, извлечённые которым куски уже присоединены к $c_{...k}$.

\subsubsection{Граф желаний и операция циклического обмена}

Основной операцией в GoLeft Protocol является обмен кусками в активных снимках между несколькими участниками по циклу. Перед этим обменом мы будем присоединять к участвующим в обмене кускам самый правый из ещё неприсоединённых ассоциированных с ним кусков, если такой кусок есть (то есть ещё не все ассоциированные куски присоединены). При обмене каждый кусок, к которому только что присоединили ассоциированный кусок, будет отдаваться тому участнику, который извлекал этот кусок. Таким образом этот участник получает в каждом активном снимке кусок, равный с его точки зрения тому куску, который он имел в начале GoLeft Protocol и инвариант 1 выполняется. Если к какому-то куску уже присоединены все ассоциированные с ним куски и есть хотя бы один участник, не извлекавший ассоциированные с ним куски, то каждый участник, не извлекавший ассоциированные с ним куски, имеет значительное преимущество над каждым участником, извлекавшим их, и никакие обмены больше не нужны. Если к какому-то куску присоединены все ассоциированные куски и все участники извлекали ассоциированные с ним куски, то этот кусок с точки зрения любого участника не меньше, чем изначальный кусок этого участника, поэтому нам не важно, кому он достанется, в любом случае инвариант 1 будет соблюдаться.

Чтобы понимать, какие куски участвуют в обмене и кто получит какие куски, будем поддерживать граф желаний.

\begin{mydef}
Граф желаний "--- орентированный граф $G = (V, E)$. $V = \{1, ..., n\}$, вершина $k$ соответствует множеству кусков $c_{...k}$, говоря о графе желаний будем отождествлять его вершины с соответствующими им множествами изоморфных кусков. $V = T \sqcup T'$, где $T'$ "--- множество вершин, к соответствующим которым кускам уже присоединены $n - 1$ ассоциированных кусков (то есть все участники извлекли в PrepareGoLeft Protocol ассоциированные с этими кусками кусоки и все эти куски уже присоединены), $T$ "--- остальные вершины. Изначально $T = \{1, ..., n\}, T' = \varnothing$. Если $k \in T'$, то в $k$ ведут рёбра из всех остальных вершин. Если $k \in T$, то в $k$ ведёт ровно одно ребро, выходящее из того множества изоморфных кусков, которым сейчас владеет участник, который извлёк самые правые из ещё неприсоединённых к $c_{...k}$ ассоциированных с кусками из $c_{...k}$ кусков.
\end{mydef}

Если к некоторому множеству кусков $c_{...k}$ из $T$ уже присоединили все ассоциированные с ним куски, то существуют участники, не извлекавшие ассоциированные с $c_{...k}$ куски и все ассоциированные с $c_{...k}$ куски уже присоединены, поэтому каждый участник, не извлекавший ассоциированные с $c_{...k}$ куски имеет значительное преимущество над каждым участником, извлекавшим ассоциированные  с $c_{...k}$ куски. В этом случае мы завершаем основную часть GoLeft Protocol, в которой используется граф желаний, и переходим к завершению протокола, в котором граф желаний не используется. Поэтому мы можем считать, что если $k \in T$, то к $c_{...k}$ присоединены ещё не все ассоциированные куски, и что каждая вершина из $T$ имеет входящую степень 1.

\begin{lemma}
В графе желаний всегда есть цикл, содержащий вершину из $T$.
\end{lemma}

\textbf{Доказательство леммы 10} Сначала покажем, что множество $T$ всегда не пусто. Действительно если $T = \varnothing$, то $T' = \{1, ..., n\}$, то есть ко всем кускам $c_{jk}$ из активных дележей было присоединено по $n - 1$ кусков. Значит в активных дележах каждый участник извлёк кусок, ассоциированный с куском каждого другого участника, то есть никто не имел ни над кем значительного преимущества. Но по утверждению 3 о свойствах Core Protocol cutter имел значительное преимущество хотя бы над одним другим участником. Получили противоречие, $\Rightarrow$ $T$ всегда не пусто. Теперь покажем, что есть цикл, содержащий хотя бы одну вершину из $T$. Возьмём произвольную вершину $v$ из $T$. Будем идти из неё по обратным рёбрам, пока не придём в уже посещённую вершину или в вершину из $T'$. Пусть $u_1, ..., u_m$ "--- получившийся путь, $u_1 = v$, $\forall i \in \{1, ..., m - 1\} (u_{i + 1}, u_i) \in E$, $u_m$ "--- первая вершина, которая уже встречалась или которая лежит в $T'$ (такой путь существует и единственен, так как все вершины из $T$ имеют входящую степень 1). Если $u_m \in T'$, то по построению $G$ в $u_m$ ведут рёбра из всех остальных вершин, поэтому $(u_1, u_m) \in E$, $(u_1 \rightarrow u_m \rightarrow u_{m - 1} \rightarrow ... \rightarrow u_2 \rightarrow u_1)$ - искомый цикл, он содержит вершину $u_1=v \in T$. Если $u_m \in T, u_m = u_k, k < m$, то $(u_m \rightarrow u_{m - 1} \rightarrow ... \rightarrow u_{k + 1} \rightarrow u_k=u_m)$ "--- искомый цикл, он содержит вершину $u_m \in T$.

Введём операцию циклического обмена. Пусть $(u_1 \rightarrow u_2  \rightarrow ... \rightarrow u_m \rightarrow u_1)$ "--- цикл в графе желаний, содержащий хотя бы одну вершину из $T$. При циклическом обмене по этому циклу сначала для всех лежащих в $T$ вершин цикла (в любом порядке) проводится присоединение к соответствующим им кускам самого правого из ещё не присоединённых кусков (при этом некоторые снимки удаляются из множества активных снимков, подробно операция присоединения ассоциированных кусков описанна в следующем разделе). Потом происходит обмен кусками в оставшихся активными снимках между участниками, владеющими кусками из цикла: участник, владевший множеством изоморфных кусков $u_m$ получает куски $u_1$, а участник, владевший $u_{i}$, получает $u_{i + 1}$, $i \in \{1, ..., m - 1\}$. После этого обновляются множества $T$ и $T'$ и перестраивается граф желаний.


\subsubsection{Операция присоединения}

Чтобы при обменах сохранялся инвариант 1, в начале операции циклического обмена мы присоединяем к участвующим в обмене множествам кусков $c_{...k}$ из $T$ самое правое из ещё не присоединённых ассоциированных с ним множеств кусков $e_{...kl}$. Мы не можем просто присоединить эти куски, так как это вызовет нарушение инварианта 2. Пусть к кускам $c_{...k}$ уже присоединены ассоциированные с ними куски $e_{...k1}, ..., e_{...kl}$ (возможно $l = 0$, тогда к этим кускам ещё не присоединены никакие куски) и мы хотим присоединить к ним куски $e_{...k(l + 1)}$ (после этого эти куски в ходе циклического обмена будут отданы участнику $i_{k(l+1)}$). Пусть $M = \{k, i_{k1}, ..., i_{kl}\}$ "--- участники, уже владевшие кусками $c_{...k}$, $M' = N \setminus M$ "--- участники, ещё не владевшие $c_{...k}$, (то есть участники, извлечённые которыми куски ещё не были присоединены к $c_{...k}$ и участники, которые не извлекали ассоциированные с $c_{...k}$ куски). Заметим, что $M$ и $M'$ не пусты, так как $k \in M$ и $i_{k(l+1)} \in M'$, $i_{k(l+1)}$ существует, так как $c_{...k} \notin T'$. Пусть $m:=|M| = l + 1$, тогда $|M'| = n - m, 1 \leq m \leq n - 1$. Перед присоединением $e_{...k(l+1)}$ к $c_{...k}$ мы выполняем следующие действия:
\begin{itemize}
  \item Участники из $M'$ по очереди (в любом порядке) выбирают по $\frac{1}{n - m + 1} \cdot |A|$ (где $A$ "--- множество активных снимков) с округлением в большую сторону ещё не выбранных другими участниками из $M'$ активных снимков. Участник $i$ выбирает активные снимки $j$ с максимальным (среди ещё не выбранных другими участниками из $M'$ активных снимков) значенинием разности $b_{ji} - (v_i(c_{jk}) + \sum_{t = 1}^l v_i(e_{jkt}))$. Если $i = i_{kl'}$ извлекал ассоциированные с $c_{...k}$ куски, то эта разность равна суммарной стоимости всех ещё неприсоединённых ассоциированных с $c_{jk}$ кусков, находящихся не левее извлечённого $i$ куска. Если $i$ не извлекал ассоциированные с $c_{...k}$ куски, то она больше, чем суммрная стоимость всех ещё не присоединённых ассоциированных с $c_{jk}$ кусков.  Все выбранные кем-то из $M'$ активные снимки удаляются из множества активных снимков. Текущий делёж в этих снимках добавляется к бонусному дележу $S_B$ (то есть к куску каждого участника в $S_B$ добавляются куски, которыми он в этот момент владел в удаляемых активных снимках, вместе со всеми уже присоединёнными к ним ассоциированными кусками). Все ещё не присоединённые куски, ассоциированные с кусками из удаляемых из $A$ снимков, возвращаются в остаток. Это действие обеспечивает то, что в инварианте 2 в правой части первое и третье слагаемые не будут рости быстрее, чем левая часть.
  \item Участники из $M$ по очереди (в любом порядке) выбирают по $\frac{m}{m^2 + 1} \cdot |A|$ с округлением в большую сторону ещё не выбранных другими участниками из $M$ снимков (мы считаем, что выбранные на прошлом шаге участниками из $M'$ снимки уже удалены из $A$). Каждый участник $i$ выбирает снимки $j$ с максимальной (среди ещё не выбранных другими участниками снимков) стоимостью присоединяемого сейчас куска $v_i(e_{jk(l+1)})$. Все выбранные снимки удаляются из множества активных снимков. Текущий делёж в этих снимках добавляется к дележу $S_B$. Все ещё не присоединённые куски, ассоциированные с кусками из удаляемых из $A$ снимков, кроме присоединяемых сейчас кусков $e_{jk(l+1)}$ возвращаются в остаток. Все куски $e_{jk(l+1)}$ из удаляемых из $A$ снимков $j$, соединяются вместе и делятся без зависти между участниками из $M$ рекурсивным запуском Main Protocol. Получившийся делёж добавляется к дележу $S_B$ Это действие обеспечивает то, что в инварианте 2 в правой части второе слагаемое не будет рости быстрее, чем левая часть.

\end{itemize}

После этих двух подготовительных действий в оставшихся после них активными снимки $j$ куски $e_{jk(l+1)}$ добавляются к $c_{jk}$.

Операция присоединения являемся атомарной для инварианта 2 то есть во время этой операции инвариант 2 может нарушаться, но вся операция целиком сохраняет этот инвариант.

\subsubsection{Реализация GoLeft Protocol}

GoLeft Protocol получает на вход множество $A$ из $C$ изоморфных снимков, в которых проведена процедура извлечения и в которых все бонусы являются очень значительными величинами или незначительными величинами и остаток $R$. GoLeft Protocol возвращает разбиение множества участников на два непустых множества $A_1$ и $A_2$ такие, что каждый участник из $A_1$ имеет значительное преимущество над каждым участником из $A_2$.

Наша реализация GoLeft Protocol использует бонусные дележи $S_{B_k}$. Они вводятся для упрощения доказательства корректности протокола, на самом деле в реализации можно не вводить эти дележи, а всё, что добавляется к ним, сразу раздавать участникам (то есть добавлять в $S_R$).

Если участник должен выбрать нецелое число снимков, подразумевается округление в большую сторону.

1. Создаём пустой бонусный дележ без зависти $S_B$ \\
2. Создаём множества $D_1, ..., D_n$, $D_k$ "--- множество участников дележа, владевших кусками $c_{...k}$. Изначально $D_k = \{k\}$. \\
3. Создаём два множества множеств изоморфных кусков $c_{...k}$ $T$ и $T'$, $T'$ состоит из тех множеств изоморфных кусков $c_{...k}$, к которым уже присоединены $n - 1$ множество ассоциированных с ними кусков $e_{...kl}$, $T$ "--- остальные множества изоморфных кусков $c_{...k}$. \\
4. While (в $T$ нет множеств кусков $c_{...k}$, к которым присоединили все ассоциированные с ними множества кусков $e_{...kl}$) do \\
5. Строим граф желаний "--- ориентированный граф $G$, в котором вершинами являются множества изоморфных кусков, в каждую вершину из $T'$ проведены рёбра из всех остальных вершин, в каждую вершину $k$ из $T$ проведено ребро из одной вершины, соответствующей тому множеству кусков, которым сейчас владеет участник, извлёкший самое правое из ещё не присоединённых к $c_{...k}$ множеств ассоциированных с кусками из $c_{...k}$ кусков $e_{...kl}$. (Такое множество существует, так как ещё не все ассоциированные с $c_{...k}$ куски присоединены, так как выполнилось условие в строке 4). \\
6. Находим в графе $G$ цикл, содержащий хотя бы одну вершину из $T$ (такой цикл есть по лемме 10). Обозначим вершины этого цикла через $u_1, ..., u_m$ (из вершины $u_i$ идёт ребро в вершину $u_{i + 1}$, из вершины $u_m$ идёт ребро в вершину $u_1$). Пусть $p_i$ "--- участник, владеющий кусками, соответствующими вершине $u_i$. \\
(Начало операции циклического обмена) \\
7. for $k \in \{u_1, ..., u_m\}$ do \\
8. if ($k \in T$) do \\
(Начало присоединения) \\
9. $l := $ число ассоциированных кусков, уже присоединённых к каждому куску из $c_{...k}$ \\
10. $M := D_k$ "--- множество участников, уже владевших кусками $c_{...k}$ (то есть $k$ и участники, чьи ассоциированные с $c_{...k}$ куски уже присоединены, $M = \{k, i_{k1}, ..., i_{kl}\}$). \\
11. $M' := N \setminus M$ "--- множество участников, не владевших $c_{...k}$ \\
12. $m := |M|$ \\
13. $A' := \emptyset$ \\
14. for $i \in M'$ do \\
15. Участник $i$ выбирает $\frac{1}{n - m + 1} \cdot |A|$ ещё не выбранных другими участниками из $M'$ активных снимков $j$ с максимальным значенинием разности $b_{ji} - (v_i(c_{jk}) + \sum_{t = 1}^l v_i(e_{jkt}))$, эти снимки добавляются в $A'$ \\
16. end for (14) \\
17. Ещё не присоединённые к кускам из $A'$ ассоциированные с ними куски возвращаются в остаток. Все снимки из $A'$ удаляются из $A$ и добавляются в $S_B$. \\
18. $A'' := \emptyset$ \\
19. for $i \in M$ do \\
20. Участник $i$ выбирает $\frac{m}{m^2 + 1} \cdot |A|$ с округлением в большую сторону ещё не выбранных другими участниками из $M$ снимков $j$ с максимальной (среди ещё не выбранных другими участниками снимков) стоимостью присоединяемого сейчас куска $v_i(e_{jk(l+1)})$. Эти снимки добавляются в $A''$. \\
21. end for (19) \\
22. Все ассоциированные с кусками из $A''$ куски $e_{jk(l + 1)}$ объединяются в один кусок $P$. Все остальные куски, ассоциированные с кусками из $A''$ и ещё не присоединённые к ним, возвращаются в остаток. \\
23. Все снимки из $A''$ удаляются из $A$ и добавляются к бонусному дележу $S_B$. \\
24. Main Protocol($P$, $M$) "--- делим кусок $P$ между участниками из $M$ рекурсивным запуском Main Protocol для меньшего числа участников. \\
25. Полученный в строке 24 делёж добавляется к дележу $S_B$ \\
26. Куски $e_{...k(l+1)}$ присоединяются к кускам $c_{...k}$ \\
(Конец присоединения) \\
end if (8) \\
end for(7) \\
27. Участники $p_1, ..., p_m$ обмениваются кусками: участник $p_i$ получает куски $c_{...u_{i+1}}$, участник $p_m$ получает куски $c_{...u_{1}}$.\\
28. Обновляем множества $D_{k}$: если множество изоморфных кусков $c_{...k}$ участвовало в обмене и лежит в $T$, получивший его участник добавляется в $D_k$.  \\
29. Обновляем множества $T$ и $T'$: если к кускам $c_{...k}$ были присоединены куски $e_{...k(n-1)}$, то $c_{...k}$ удаляется из $T$ и добавляется в $T'$. \\
(Конец операции циклического обмена) \\
30. end while (4) \\
31. Пусть $c_{...k}$ "--- множество кусков из $T$, к которому присоединили все ассоциированные с ними множества кусков $e_{...kl}$. \\
32. $A_1 := N \setminus D_k$ \\
33. $A_2 := D_k$ \\ 
34. Добавляем к $S_R$ делёж $S_B$ и текущие дележи во всех активных снимках, все ещё не присоединённые ассоциированные куски возвращаются в остаток. \\
35. return $(A_1, A_2)$ \\

\subsubsection{Корректность GoLeft Protocol}

\begin{lemma}
Инвариант 1 выполнен в начале GoLeft Protocol и после каждой операции циклического обмена.
\end{lemma}

\textbf{Доказательство леммы 11} В начале GoLeft Protocol каждый участник $k$ имеет в каждом активном дележе $j$ ровно $b_{jk}$ (по определению $b_{jk}$). Если участник не участвовал в циклическом обмене, то во всех оставшихся активными дележах его кусок не изменился. Если участник $k$ участвовал в циклическом обмене и получил куски из $T$, то к куску из каждого оставшегося активным дележа $j$ в этом циклическом обмене был присоединён извлечённый этим участником кусок, до этого были пресоединены все более правые ассоциированные куски, поэтому по мнению $k$ стоимость этого куска равна $b_{jk}$. Если участник $k$ участвовал в циклическом обмене и получил куски из $T'$, то к куску из каждого оставшегося активным дележа $j$ уже присоединены все ассоциированные с ним куски, в том числе кусок, извлечёный участником $k$, и все более правые куски, поэтому его стоимость с точки зрения $j$ хотя бы $b_{jk}$.

\begin{lemma}
Инвариант 2 выполнен в начале GoLeft Protocol.
\end{lemma}

\textbf{Доказательство леммы 12} В начале GoLeft Protocol никакие куски ещё не присоединены (то есть для всех $k$ величина $g_k$ из формулировки инварианта равна 0), поэтому ни для каких $i$ и $i'$ нет таких $k$, что $l'_k < g_k$ или $l_k < g_k$, поэтому множества $M_1$, $M_2$ и $M_3$ пусты, все $bv_{ii'}$ равны 0, поэтому инвариант выполняется.

\begin{lemma}
Инвариант 2 сохраняется при удалении некоторых снимков из множества активных снимков и их добавлении в $S_B$.
\end{lemma}

\textbf{Доказательство леммы 13} Достаточно доказать для удаления одного снимка, утверждение для произвольного числа снимков получаем по индукции, удаляя снимки по очереди. Пусть $j$ "--- удаляемый снимок. Рассмотрим участников дележа $i$ и $i'$.  Пусть $i'$ владеет в этот момент куском $c_{jk}$. По мнению $i$ в дележе, добавляемом к $S_B$, $i$ получает хотя бы $b_{ji}$, а $i'$ получает ровно $v_i(c_{jk}) + \sum_{t = 1}^{g_k} v_i(e_{jkt})$. Если $i$ не извлекал ассоциированный с $c_{jk}$ кусок или этот кусок ещё не был присоединён, то $b_{ji} \geq v_i(c_{jk}) + \sum_{t = 1}^{g_k} v_i(e_{jkt})$, левая часть в неравенстве не уменьшается, а правая часть не увеличивается, поэтому неравенство сохраняется. Если $i$ извлекал ассоциированный с $c_{jk}$ кусок и он уже был присоединён, то $b_{ji} = v_i(c_{jk}) + \sum_{t = 1}^{l_k} v_i(e_{jkt})$, левая часть неревенства уменьшается на $\sum_{t = l_k + 1}^{g_k} v_i(e_{jkt})$, а правая часть уменьшается на $\sum_{k' \in M_1} \sum_{t = g_{k'} + 1}^{l_{k'}} v_i(e_{jk't}) + \sum_{k' \in M_2} \sum_{t = l_{k'} + 1}^{g_{k'}} v_i(e_{jk't}) + \sum_{k' \in M_3} (\sum_{t = g_{k'} + 1}^{m_{k'}} v_i(e_{jk't}) + \frac {1}{nC'}v_i(R))$, то есть на большую величину, так как $k \in M_2$.

Для доказательства того, что инвариант 2 сохраняется при операциях присоединения разберём разные расположения извлечённых $i$ и $i'$ кусков относительно присоединяемого куска и для каждого их расположения покажем, что для $i$ и $i'$ инвариант сохраняется. В леммах 17, 18, 19 рассмотренны нетривиальные случаи, для которых нам потребовались удаления снимков из остатка в строках 13-17 и 18-23 алгоритма. Остальные случаеи достаточно тривиальны, однако мы приведём аккуратные доказательства и для них в леммах 14, 15, 16.

\begin{lemma}
Пусть проводится операция присоединения кусков $e_{...k(g_k + 1)}$ к куску $c_{...k}$, $i$ и $i'$ таковы, что и $i$ и $i'$ не извлекали ассоциированные с $c_{...k}$ куски или извлекали их, но они ещё не были присоединены (но возможно присоединяются сейчас) к $c_{...k}$ (то есть $l_k > g_k$ или $l_k$ не определено и $l_k' > g_k$ или $l_k'$ не определено). Тогда инвариант 2 для $i$ и $i'$ сохраняется при операции присоединения.
\end{lemma}

\textbf{Доказательство леммы 14} Инвариант 2 сохраняется при удалении снимков, никакие другие операции не меняют входящие в условие из инварианта 2 величины: $i$ и $i'$ не участвуют в дележе в строке 24, присоединение кусков в строке 26 не влияет на условие из инварианта, так как для $i$ и $i'$ $k$ не лежит ни в одном из множеств $M_1$, $M_2$, $M_3$.

\begin{lemma}
Пусть проводится операция присоединения кусков $e_{...k(g_k + 1)}$ к куску $c_{...k}$, $i$ и $i'$ таковы, что и $i$ и $i'$ извлекали ассоциированные с $c_{...k}$ куски и они уже были присоединены (до этого присоединения),то есть $l_k \leq g_k$ и $l_k' \leq g_k$. Тогда инвариант 2 для $i$ и $i'$ сохраняется при операции присоединения.
\end{lemma}

\textbf{Доказательство леммы 15} Инвариант 2 сохраняется при удалении снимков из множества активных снимков. При дележе в строках 24-25 инвариант сохраняется, так как правая часть неравенства не меняется, а левая часть не уменьшается, так как в $S_B$ добавляется делёж без зависти, в котором участвуют и $i$, и $i'$. Если $i'$ не владеет в начале присоединения кусками $c_{...k}$, то присоединение кусков в строке 26 не влияет на условие из инварианта, так как для $i$ и $i'$ $k$ не лежит ни в одном из множеств $M_1$, $M_2$, $M_3$, а если $i'$ владеет в начале присоединения кусками $c_{...k}$, то присоединение кусков в строке 26 уменьшает правую часть, так как $k$ перестаёт лежать в $M_2$.

\begin{lemma}
Пусть проводится операция присоединения кусков $e_{...k(g_k + 1)}$ к куску $c_{...k}$. Пусть $i$ не извлекал куски, ассоциированные с $c_{...k}$, или извлекал куски, ассоциированные с $c_{...k}$, но они ещё не были присоединёны. Пусть $i'$ извлекал кусок, ассоциированный с $c_{...k}$, который уже был присоединён и $i'$ сейчас не владеет $c_{...k}$. Тогда инваринт 2 для $i$ и $i'$ сохраняется при операции присоединения.
\end{lemma}

\textbf{Доказательство леммы 16} Инвариант сохраняется при удалении снимков из $A'$ в строке 17 и при присоединении в строке 26.
Покажем, что он сохраняется при удалении снимков из $A''$ и последующем дележе в строках 22-25. В этом дележе $i$ не участвует, а $i'$ участвует и получает кусок от $\bigcup_{j \in A''} e_{jk(g_k+1)}$, левая часть уменьшается не более, чем на $\sum_{j \in A''} v_i(e_{jk(g_k+1)})$. При удалении снимков из $A''$ левая часть неревенства не уменьшается или, если $i'$ владеет сейчас кусками $c_{...k''}, k'' \in M_2$, уменьшается не более, чем на $\sum_{j \in A''} \sum_{t = l_{k''} + 1}^{g_{k''}} v_i(e_{jk''t})$, а правая часть уменьшается на $\sum_{j \in A''} (\sum_{k' \in M_1} \sum_{t = g_{k'} + 1}^{l_{k'}} v_i(e_{jk't}) + \sum_{k' \in M_2} \sum_{t = l_{k'} + 1}^{g_{k'}} v_i(e_{jk't}) + \sum_{k' \in M_3} (\sum_{t = g_{k'} + 1}^{m_{k'}} v_i(e_{jk't}) + \frac {1}{nC'}v_i(R))) $, поэтому правая часть уменьшается хотя бы на \\ $\sum_{j \in A''} (\sum_{k' \in M_1} \sum_{t = g_{k'} + 1}^{l_{k'}} v_i(e_{jk't}) + \sum_{k' \in M_3} (\sum_{t = g_{k'} + 1}^{m_{k'}} v_i(e_{jk't}) + \frac {1}{nC'}v_i(R)) ) $ больше, чем левая часть. По условию леммы для $i$ и $i'$ $k \in M_1 \cup M_3$, поэтому правая часть в строках 22-23 уменьшается хотя бы на $\sum_{j \in A''} v_i(e_{jk(g_k+1)})$ больше, чем левая часть, поэтому в строках 22-25 левая часть неравенства уменьшается не больше, чем правая часть, поэтому неравенство остаётся верным. \\
Неформально инвариант гарантировал нам, что $i$ не будет завидовать $i'$ даже если отдать $i'$ все ещё не присоединённые к $c_{...k}$ куски, расположенные правее извлечённого $i$ куска, после того как мы удалили из множества не присоединённых к $c_{...k}$ кусков куски из $A''$ и часть из них отдали $i'$, инвариант продолжил выполняться.

\begin{lemma}
Пусть проводится операция присоединения кусков $e_{...k(g_k + 1)}$ к куску $c_{...k}$. Пусть $i$ извлекал кусок, ассоциированный с $c_{...k}$, который уже был присоединён (и возможно всё ещё владеет $c_{...k}$). Пусть $i'$ не извлекал куски, ассоциированные с $c_{...k}$, или извлекал куски, ассоциированные с $c_{...k}$, но они ещё не были присоединёны. Тогда инваринт 2 для $i$ и $i'$ сохраняется при операции присоединения.
\end{lemma}

\textbf{Доказательство леммы 17} Инвариант сохраняется при удалении снимков. Пусть $A''$ "--- снимки, выбранные в строках 19-21, $A$ "--- снимки, оставшиеся активными после этой операции присоединения. Тогда $i$ выбрал в множество $A''$ хотя бы в $m$ раз больше снимков, чем осталось в $A$ (где $m$ "--- количество участников, уже владевших $c_{...k}$), так как $\frac {m}{m^2 + 1} = m * (1 - \frac {m}{m^2 + 1})$, от того что при делении мы округляем в большую сторону левая часть не уменьшилась, а правая не увеличилась. В каждом взятом $i$ снимке $j$ по мнению $i$ кусок $e_{...k(g_k+1)}$ не меньше, чем в каждом снимке из $A$, поэтому стоимость $P$ по мнению $i$ хотя бы в $m$ раз больше, чем суммарная стоимость кусков $e_{...k(g_k + 1)}$, оставшихся в $A$. При дележе в строках 24-25 $i$ участвует в дележе, а $i'$ не участвует, из отсутсвия зависти следует пропорциональность, поэтому $bv_{ii'}$ увеличивается хотя бы на $\frac{v_i(P)}{m}$, то есть хотя бы на суммарную стоимость кусков $e_{...k(g_k + 1)}$, оставшихся в $A$, а правая часть неравенства не меняется. В строке 26 левая часть не меняется, а правая часть увеличивается на суммарную стоимость кусков $e_{...k(g_k + 1)}$, оставшихся в $A$, поэтому всего в строках 24-26 левая часть неравенства увеличится больше, чем правая, поэтому неравенство останется верным.

\begin{lemma}
Пусть проводится операция присоединения кусков $e_{...k(g_k + 1)}$ к куску $c_{...k}$. Пусть $i$ извлекал куски, ассоциированные с $c_{...k}$, но они ещё не были присоединёны. Пусть $i'$ владеет $c_{...k}$ в начале операции присоединения. Тогда инваринт 2 для $i$ и $i'$ сохраняется при операции присоединения.
\end{lemma}

\textbf{Доказательство леммы 18} В начале операции присоединения $i$ имеет в каждом активном снимке $j$ кусок, не меньший чем $b_{ji}$ (по инварианту 1), поэтому бонус $i$ над $i'$ в этом снимке, не меньше, чем $b_{ji} - (v_i(c_{...k}) + \sum_{t = 1}^{g_k} v_i(e_{jkt}) = \sum_{t = g_k + 1}^{l_k} v_i(e_{jkt})$ В строках 14-16 $i$ выбирает и добавляет в $A'$ не меньше снимков, чем останется в $A$ после удаления из него выбранных в строках 14-16 снимков, в каждом из выбранных $i$ снимков $\sum_{t = g_k + 1}^{l_k} v_i(e_{jkt})$ не меньше, чем в каждом из оставшихся в $A$, поэтому в строках 14-17 разность между левой и правой частями неравенства увеличится хотя бы на $\sum_{j \in A'} \sum_{t = g_k + 1}^{l_k} v_i(e_{jkt}) \geq \sum_{j \in A} \sum_{t = g_k + 1}^{l_k} v_i(e_{jkt})$ (где $A$ "--- множество оставшихся активными после строки 17 снимков), то есть не меньше, чем правая часть неравенства уменьшается при добавлении $k$ в множество $M_1$, которое произойдёт в результате этого присоединения. Таким образом уменьшение множества активных снимков вместе с добавлением $k$ в множества $M_1$ сохраняют инвариант. Уменьшение множества активных снимков и делёж в строках 22-25 также сохраняют инвариант, доказательство этого совпадает с доказательством леммы 16 (мы считаем, что $k$ добавили в $M_1$ сразу после строк 14-17, в этом доказательстве считаем, что $k$ уже лежит в $M_1$). \\
Неформально инвариант гарантировал, что $i$ не будет завидовать никому из тех, кто уже владел кусками $c_{...k}$ и уже отдал их, даже если отдать ему все ещё не присоединённые к $c_{...k}$ куски, расположенные правее извлечённого $i$ куска. Кроме того, в каждом активном снимке $j$ $i$ имеет сейчас бонус над текущим владельцем $c_{...k}$, не меньший, чем суммарная стоимость всех ещё не присоединённых к $c_{jk}$ кусков, расположенных правее извлечённого им куска. Зафиксировав текущий делёж в достаточно большой доле активных снимков, выбранных так, чтобы эта суммарная стоимость была максимальной, и удалив их из множества активных дележей, мы добиваемся того, что $i$ не будет завидовать и текущему владельцу, если отдать ему все ещё не присоединённые к $c_{...k}$ куски, расположенные правее извлечённого $i$ куска.

\begin{lemma}
Пусть проводится операция присоединения кусков $e_{...k(g_k + 1)}$ к куску $c_{...k}$. Пусть $i$ не извлекал куски, ассоциированные с $c_{...k}$. Пусть $i'$ владеет $c_{...k}$ в начале операции присоединения. Тогда инваринт 2 для $i$ и $i'$ сохраняется при операции присоединения.
\end{lemma}

\textbf{Доказательство леммы 19} В начале операции присоединения $i$ имеет в каждом активном снимке $j$ кусок, не меньший чем $b_{ji}$ (по инварианту 1), поэтому бонус $i$ над $i'$ в этом снимке, не меньше, чем $b_{ji} - (v_i(c_{...k}) + \sum_{t = 1}^{g_k} v_i(e_{jkt})$. $i$ не извлекал ассоциированные с $c_{...k}$ куски, значит он считал, что имеет в активных снимках значительное преимущество над $k$, а значит и очень значительное преимущество над $k$ (так как в PrepareGoLeft Protocol мы добились того, чтобы в каждом снимке каждый участник имел над каждым другим участников очень значительное преимущество или незначительное преимущество), то есть для любого активного дележа $j$ $b_{ji} - v_i(c_{jk}) \geq \frac{1}{C'} \cdot v_i(R)$. Все участники считают все извлечённые куски незначительными, поэтому $b_{ji} - (v_i(c_{jk}) + \sum_{t = 1}^{m_k} v_i(e_{jkt})) \geq \frac{1}{C'} \cdot v_i(R) - m \cdot \frac{1}{n^2C'} \cdot v_i(R) \geq \frac{1}{nC'} \cdot v_i(R)$, то есть $b_{ji} - (v_i(c_{...k}) + \sum_{t = 1}^{g_k} v_i(e_{jkt})) \geq \sum_{t=g_k+1}^{m_k}v_i(e_{jkt}) + \frac{1}{nC'} \cdot v_i(R)$. Дальше так же как в лемме 18 можно доказать, что уменьшение множества активных снимков в строках 14-17 и добавление $k$ в множество $M_3$ для $i$ и $i'$ сохраняют инвариант. То, что уменьшение множества активных снимков  и делёж в строках 22-25 сохраняют инвариант доказывается так же, как лемма 16.

\begin{lemma}
Если инварианты 1 и 2 выполняются, для какого-то $k$ из $T$ все ассоциированные с $c_{...k}$ куски были присоединены к $c_{...k}$ и $|A| \geq 1$, то делёж $S_A \cup S_B$ является частичным дележём без зависти, причём каждый участник из $D_k$ имеет в нём значительное преимущество над каждым участником из $N \setminus D_k$, где $D_k$ "--- множество участников дележа, владевших кусками $c_{...k}$.
\end{lemma}

\textbf{Доказательство леммы 20} \\
1. Покажем, что в дележе $S_A \cup S_B$ никакой участник $i$ не завидует никакому участнику $i'$. Пусть $i'$ владеет кусками $c_{...k'}$. Если $i$ ещё не владел кусками $c_{...k'}$ (то есть он не извлекал ассоциированные с $c_{...k'}$ куски или извлекал, но они ещё не были присоединены), то $i$ не завидует $i'$ ни в $S_A$, ни в $S_B$, поэтому $i$ не завидует $i'$ в $S_A \cup S_B$. Если извлечённые $i$ куски уже присоедины к $c_{...k'}$, то для $i$ и $i'$ $k' \in M_2$, поэтому по инварианту 2 $bv_{ii'} \geq \sum_{j \in A} \sum_{t = l_k' + 1}^{g_k'} v_i(e_{jk't})$, в каждом активном дележе $j$ $i'$ имеет по мнению $i$ $b_{ji} + \sum_{t = l_k' + 1}^{g_k'} v_i(e_{jk't})$, а $i$ "--- не меньше, чем $b_{ji}$ (по инварианту 1), поэтому $i$ считает, что в дележе $S_A$ $i'$ имеет не более, чем на $\sum_{j \in A} \sum_{t = l_k' + 1}^{g_k'} v_i(e_{jk't})$ больший кусок, чем $i$, поэтому в дележе $S_A \cup S_B$ $i$ не завидует $i'$. \\
2. Покажем, что если $i \in N \setminus D_k$, $i' \in D_k$, то в дележе $S_A \cup S_B$ $i$ имеет значительное преимущество над $i'$. Если $i'$ сейчас владеет $c_{...k}$, то $i$ имеет значительное преимущество над $i'$ в каждом активном дележе (так как $i$ имеет в активном дележе $j$ хотя бы $b_{ji}$, а $i'$ имеет по мнению $i$ $v_i(c_{jk}) + \sum_{t = 1}^{m_k} v_i(e_{jkt}) \leq b_{ji} - \frac{1}{C'}v_i(R) + m_k \cdot \frac{1}{n^2C'}v_i(R) \leq b_{ji} - \frac{1}{nC'}v_i(R)$), $S_B$ является дележём без зависти (так как по инварианту 2 все бонусы неотрицательны), поэтому $i$ имеет над $j$ значительное преимущество в $S_A \cup S_B$. Если $i'$ уже не владеет $c_{...k}$, то пусть $i'$ владеет $c_{...k'}$. Тогда для $i$ и $i'$ $k \in M_3$. Если $i$ не владел $c_{...k'}$, то $i$ не завидует $i'$ в $S_A$ и имеет по инварианту 2 бонус над $i'$ в $S_B$ не меньшее, чем $|A| \cdot \frac{1}{nC'}v_i(R) \geq \frac{1}{nC'}v_i(R)$, поэтому $i$ имеет над $i'$ значительное преимущество в $S_A \cup S_B$. Если $i$ владел $c_{...k'}$, то $k' \in M_2$, поэтому по инварианту 2 $bv_{ii'} \geq \sum_{j \in A} \sum_{t = l_{k'} + 1}^{g_{k'}} v_i(e_{jk't}) + |A| \cdot \frac{1}{nC'}v_i(R)$, поэтому бонус $i$ над $i'$ в $S_A \cup S_B$ не меньше, чем $|A| \cdot \frac{1}{nC'}v_i(R)$, поэтому $i$ имеет значительное преимущество над $i'$ в $S_A \cup S_B$.

Из лемм 14-20 следует следующая теорема о корректности GoLeft Protocol:

\begin{theorem}
Если MainProtocol для любого меньшего, чем $n$ числа участников строит делёж без зависти, и при выполнении GoLeft Protocol для $n$ участников множество активных снимков не становится пустым, то GoLeft Protocol для $n$ участников возвращает разбиение множества участников на два непустых множества $A_1$ и $A_2$ такие, что каждый участник из $A_1$ имеет значительное преимущество над каждым участником из $A_2$ (в текущем частичном дележе без зависти $S_R$).
\end{theorem}

Значение константы $C$ важно только для того, чтобы множество активных снимков не стало пустым. Можно взять любую такую константу $C$, что множество активных снимков не станет пустым, но при увеличении $C$ будет увеличиваться время работы протокола. Поэтому то, что $C = n^4 \cdot 2^{n^2} \cdot n^{3n^2}$ подходит мы доказываем в следующем разделе, в котором оценивается время работы GoLeft Protocol.

\textbf{Доказательство теоремы 3} По лемме 12 в начале GoLeft Protocol выполнен инвариант 2, по леммам 14-19 он сохраняется при операциях присоединения, никакие другие операции не влияют на него, поэтому он верен в конце GoLeft Protocol. По лемме 11 выполняется инвариант 1. По лемме 20 из этого следует, что в конце работы GoLeft Protocol каждый участник из $A_1$ имеет значительное преимущество над каждым участником из $A_2$ в дележе $S_A \cup S_B$, в строке 34 этот делёж добавляется к частичному дележу без зависти $S_R$, поэтому после этого каждый участник из $A_1$ имеет значительное преимущество над каждым участником из $A_2$ в дележе $S_R$.

\subsubsection{Оценка времени работы GoLeft Protocol}

Как было обещано выше докажем, что $C = n^4 \cdot 2^{n^2} \cdot n^{3n^2}$ подходит.

\begin{lemma}
При выполнении GoLeft Protocol количество активных снимков не становится меньше, чем $n^4$.
\end{lemma}

\textbf{Доказательство леммы 21} Операция присоединения запускается не более, чем $n(n-1) < n^2$ раз (так как к каждому из $n$ множеств кусков $c_{...k}$ присоединяется не более, чем $n - 1$ множеств кусков $e_{...kl}$). При каждой операции присоединения в строке 17 из $A$ удаляется не более, чем $(n - m) \cdot (\frac{1}{n - m + 1} \cdot |A| + 1)$ снимков ($1$ возникает из-за округления в большую сторону), поэтому при $|A| > 2n^2$ остаётся хотя бы $\frac{1}{n-m+1} \cdot |A| - (n - m) \geq \frac{1}{n} \cdot |A| - n \geq \frac{1}{2n} \cdot |A|$ активных снимков, то есть $|A|$ уменьшается не более, чем в $2n$ раз; в строке 23 из $A$ удаляется не более, чем $m \cdot (\frac{m}{m^2 + 1} \cdot |A| + 1)$ активных снимков, при $|A| > n^4$ остаётся хотя бы $\frac{1}{m^2 + 1} \cdot |A| - m \geq \frac{1}{n^2 - 2n + 2} \cdot |A| - n = \frac{|A|}{n^2} + \frac{2n - 2}{n^2(n^2 - 2n + 2)} \cdot |A| - n \geq \frac{|A|}{n^2} + (\frac{|A|}{n^3} - n) \geq \frac{|A|}{n^2}$, то есть $|A|$ уменьшается не более, чем в $n^2$ раз. Значит всего за одну операцию присоединения $|A|$ уменьшается не более, чем в $2n^3$ раз, значит за всю работу GoLeft Protocol $A$ уменьшается не более, чем в $(2n^3)^{n^2} = 2^{n^2} \cdot n^{3n^2}$ раз, то есть не становится меньше, чем $n^4$.

Оценим число запросов в GoLeft Protocol.

\begin{theorem}
GoLeft Protocol для $n$ участников не более, чем $n(n - 1)$ раз запускает Main Protocol для меньшего числа участников и больше не делает никаких запросов присоединения.
\end{theorem}

\textbf{Доказательство теоремы 4} К каждому из $n$ множеств кусков $c_{...k}$ присоединяется не более, чем $n - 1$ множество кусков $e_{...kl}$, при каждом присоединении 1 раз вызывается Main Protocol для меньшего количества участников, кроме этого GoLeft Protocol не делает никаких операций присоединения.

 	\subsection{Main protocol}

Main Protocol делит пирог без зависти и без остатка между всеми участниками. При $n \leq 3$ Main Protocol запускает один из ранее известных протоколов дележа без зависти, работающих за ограниченное число операций. При $n \geq 4$ Main Protocol запускает PrepareGoLeft Protocol и GoLeft Protocol, GoLeft Protocol возвращает разбиение множества участников $N$ на два непустых подмножества $A_1$ и $A_2$, таких, что каждый участник из $A_1$ имеет значительное преимущество над каждым участником из $A_2$. После этого с помощью $B$ запусков Core Protocol с каждым из участников из $A_1$ в кажестве $cutter$ Main Protocol добивается того, что каждый участник из $A_1$ доминирует над каждым участником из $A_2$, после чего участники из $A_1$ исключаются из дележа и рекурсивно запускается Main Protocol с меньшим числом участников для дележа остатка между участниками из $A_2$.

\subsubsection{Реализация Main Protocol}

Main Protocol принимает на вход кусок пирога $R$ и множество участников дележа $N$ и возвращает делёж $S_R$ куска $R$ между участниками из $N$ без зависти и без остатка.

1. if $|N| \leq 3$ then \\
2. $R$ делится без зависти и без остатка между участниками из $N$ с помощью одного из ранее известных ограниченных протоколов дележа без зависти. \\
3. else (1) \\
4. Созвать пустой делёж $S_R$ \\
5. Запустить PrepareGoLeft Protocol($N$, $R$). При всех запусках Core Protocol, кроме запусков в строке 1 протокала, построенный делёж добавляется к $S_R$ \\
6. if PrepareGoLeft Protocol вернул $a = 0$ then \\
7. Пусть PrepareGoLeft Protocol вернул множество изоморфных снимков $A$ и остаток $R$ \\
8. $(A_1, A_2) = $ GoLeft($A$, $R$) \\
9. for $i \in A_1$ do \\
10. $B$ раз запустить Core Protocol ($cutter=i$), добавить возвращаемые ими дележи к $S_R$ \\
11. end for (9) \\
12. Пусть $R$ "--- текущий остаток \\
13. Запустить Main Protocol($R$, $A_2$), добавить возвращаемый им делёж в $S_R$ \\
14. end if (6) \\
15. return $S_R$ \\
16. end if (1) \\

\subsubsection{Корректность и оценка времени работы Main Protocol}

Из ранее доказанных теорем о корректности и времени работы PrepareGoLeft Protocol и GoLeft Protocol и свойств Core Protocol нетрудно вывести следующую теорему, являющуюся основным результатом этой работы.

\begin{theorem}
Main Protocol строит делёж без зависти и без остатка и использует не более, чем $n^{8n^2(1 + o(1))}$ операций.
\end{theorem}

\textbf{Доказательство теоремы 5} \\
1. Докажем корректность индукцией по количеству участников $n$. База для $n \leq 3$ следует из корректности протоколов для двух и трёх участников. Из предположения индукции и теоремы 1 получаем, что в строке 5 PrepareGoLeft Protocol строит делёж без зависти без остатка и возвращает $a = 1$ (тогда Main Protocol завершается) или строит частичный делёж без зависти и возвращает $a = 0$, множество из $C$ снимков $A$, удовлетворяющих требованиям к входным данным для GoLeft Protocol и остаток $R$. Из предположения индукции, леммы 21 и теоремы 3 следует, что в строке 8 GoLeft Protocol возвращает разбиение множества участников на два непустых множества $A_1$ и $A_2$ такие, что каждый участник из $A_1$ имеет значительное преимущество над каждым участником из $A_2$ в дележе $S_R$. По утверждению 4 о свойствах Core Protocol после запусков Core Protocol в строках 9-11 каждый участник из $A_1$ будет доминировать над каждым участником из $A_2$. Поэтому после дележа остатка между участниками из $A_2$ в строке 13 (который по предположению индукции является дележом без зависти) никто не будет никому завидовать. \\
2. Оценим время работы Main Protocol. При $n \geq 3$ Main Protocol для $n$ участников один раз запускает PrepareGoLeft Protocol для $n$ участников в строке 5, не более одного раза (если PrepareGoLeft Protocol не вернул $a = 1$) запускает GoLeft Protocol для $n$ участников в строке 8, не более чем $nB$ раз запускает Core Protocol в строке 10, не более одного раза запускает Main Protocol в строке 13 и больше не делает никаких операций присоединения. Применяя теорему 2' о количестве операций в PrepareGoLeft Protocol, теорему 4 о количестве операций в GoLeftProtocol и лемму 1 о количестве операций в Core Protocol и подставляя $B = O(n^4)$, получаем, что Main Protocol не более, чем $2 + n(n - 1) + 1 \leq n^2$ (так как $n > 3$) раз запускает Main Protocol для меньшего числа участников и кроме этого делает не более, чем $n^{8n^2(1 + o(1))} + n \cdot O(n^4) \cdot {(\frac{3 + \sqrt{5}}{2})}^n = n^{8n^2(1 + o(1))}$ операций разрезания. Дерево рекурсии для запусков Core Protocol имеет глубину не больше, чем $n$, и каждая вершина имеет не более, чем $n^2$, детей, поэтому на глубине $k$ оно имеет не более, чем $(n^2)^k$ вершин, всего не более, чем $(n^2)^{n - 3} \leq n^{2n}$ запусков Main Protocol($n'$) для $n' \leq 3$ и не более, чем $\sum_{t = 0}^{n - 4} (n^2)^t \leq (n^2)^{n - 4} \cdot (n - 3) \leq n^{2n}$ запусков Main Protocol($n'$) для $3 < n' \leq n$. Оценив количество операций разрезания в Main Protocol($n'$) вне других рекурсивных вызовов Main Protocol для $n' \leq 3$ константой $X$, зависящей только от выбранного в строке 2 протокола, а для $n' \in \{4, ..., n\}$ величиной $n^{8n^2(1 + o(1))}$ (то есть оценкой для максимального возможного $n'$), получаем, что всего Main Protocol($n$) выполняет не более, чем $n^{2n} \cdot X + n^{2n} \cdot n^{8n^2(1 + o(1))} = n^{8n^2(1 + o(1))}$ операций разрезания. По замечанию 1 Main Protocol выполняет не более, чем $(n - 1) \cdot n^{8n^2(1 + o(1))} + n = n^{8n^2(1 + o(1))}$ операций измерения, то есть всего не более, чем $n^{8n^2(1 + o(1))} + n^{8n^2(1 + o(1))} = n^{8n^2(1 + o(1))}$ операций.

\end{document}